\documentclass[a4paper,11pt]{article}
\pdfoutput=1 
\usepackage{jcappub}
\usepackage{changepage}
\usepackage{amsmath,amssymb,upgreek}
 \usepackage{xcolor}
\usepackage{soul}
 \usepackage{multirow}
\usepackage[toc,page]{appendix}
 \usepackage{appendix}
\usepackage{setspace}
\usepackage{subcaption}
\usepackage{bm}
\usepackage{fontawesome5}
\usepackage{cleveref}
\usepackage[normalem]{ulem}
\newcommand{\github}[1]{%
   \href{#1}{\faGithubSquare}%
}
\makeatother

\usepackage{siunitx}
\usepackage{booktabs}

\newcommand{\Mc}{\mathcal{M}_{\rm c}}

\title{Detectability and Parameter Estimation for Einstein Telescope Configurations with \texttt{GWJulia}}

\author[a,b]{Andrea Begnoni,}
\author[b,a,c]{Stefano Anselmi,}
\author[d,e]{Mauro Pieroni,}
\author[a,b]{Alessandro Renzi,}
\author[f,g]{Angelo Ricciardone}

\affiliation[a]{Dipartimento di Fisica e Astronomia ``G. Galilei'', Universit\`a degli Studi di Padova, via Marzolo 8, I-35131 Padova, Italy}
\affiliation[b]{INFN Sezione di Padova, I-35131 Padova, Italy}
\affiliation[c]{LUTH, UMR 8102 CNRS, Observatoire de Paris, PSL Research University, Universit\'e Paris Diderot, 92190 Meudon, France}
\affiliation[d]{Instituto de Estructura de la Materia (IEM), CSIC, Serrano 121, 28006 Madrid, Spain}
\affiliation[e]{CERN, Theoretical Physics Department, Esplanade des Particules 1, Geneva 1211, Switzerland}
\affiliation[f]{Dipartimento di Fisica “Enrico Fermi”, Università di Pisa,
Largo Bruno Pontecorvo 3, Pisa I-56127, Italy}
\affiliation[g]{
INFN, Sezione di Pisa,
Largo Bruno Pontecorvo 3, Pisa I-56127, Italy}

\emailAdd{andrea.begnoni@phd.unipd.it}
\emailAdd{stefano.anselmi@pd.infn.it}
\emailAdd{mauro.pieroni@csic.es}
\emailAdd{alessandro.renzi@pd.infn.it}
\emailAdd{angelo.ricciardone@unipi.it}

\subheader{{\rm CERN-TH-2025-xxx}}

\abstract{Future gravitational-wave (GW) detectors are expected to detect tens of thousands of compact binary coalescences (CBC) per year, depending also on the final detectors layout. For this reason, it is essential to have a fast, reliable tool for forecasting how different detector layouts will affect parameter estimation for these events. The Fisher Information Matrix (FIM) is a common tool for tackling this problem. In this paper, we present a new open source code \texttt{GWJulia} to perform FIM analysis of CBC parameters, i.e., stellar black-hole binaries (BBH), neutron star binaries (BNS), and neutron star-black hole binaries (NSBH). The code is purely written in Julia, making it fast while maintaining a high level of accuracy. We consider a set of case studies to compare different Einstein Telescope (ET) designs. We compare a 10km triangular configuration with two 15km L-shaped detectors with different orientations and temperatures. We discuss also the accuracy of combinations of parameters, which is very informative for cosmology or population studies. Finally, we focus on the detection of golden events and explore how the FIM can guide posterior sampling of GW signals using a novel Hamiltonian Monte Carlo (HMC) sampler. The code is publicly available~\github{https://github.com/andrea-begnoni/GW.jl}}

\arxivnumber{2506.21530}

\NewDocumentCommand{\evalat}{sO{\big}mm}{%
  \IfBooleanTF{#1}
   {\mleft. #3 \mright|_{#4}}
   {#3#2|_{#4}}%
}
\begin{document}

\maketitle

\section{Introduction}

    Thanks to the groundbreaking detections made by the LIGO-Virgo-KAGRA Collaboration (LVK)~\cite{LIGOScientific:2016_150914, LIGOScientific:2017_0817, KAGRA:2021TC3, LIGOScientific:2017MM}, the field of Gravitational Wave (GW) has experienced a remarkable growth in recent years. After three observing runs, the catalog of GW events is close to 90 binary black hole (BBH) mergers, detected at a rate of roughly one per week, in addition to two Binary Neutron Star (BNS) mergers and two Neutron Star–Black Hole (NSBH) events. All these events had unexpected implications for astrophysics, cosmology, and fundamental physics~\cite{LIGOScientific:2021GR, KAGRA:2021pop}. Moreover, these measurements brought the community to push forward the development of third-generation interferometers. In space, we will have the LISA detector~\cite{LISA:2017amaro}, a giant equilateral interferometer to probe the GW sky in the milli-Hz frequency band with the possibility to shed light on new astrophysical~\cite{LISA:2022astro} and cosmological sources~\cite{LISACosmologyWorkingGroup:2022cosmo}. On the ground, the GW community is working on the development of the Einstein Telescope (ET)~\cite{ET:2019Magg, Abac:2025BB}, an ambitious detector aimed at detecting signals up to very high redshifts, including sources from primordial epochs. On the US side, the GW community is pursuing a parallel path with the Cosmic Explorer (CE)~\cite{Evans:2021CE, Reitze:2019CE}. These next-generation detectors — LISA, ET, and CE — will achieve at least a factor of ten improvement in the sensitivity and expand the frequency range. The final layout for the ground-based detectors has not been finalized. Several case study analyses are ongoing to compare the impact of different configurations on astrophysical and cosmological GW aspects~\cite{Abac:2025BB, Maggiore:2024_3G, Iacovelli:2024mjy}. For ET, there are two proposals for the configuration: a triangular 10km underground configuration with the detector placed in Sardinia~\cite{2020JGRB..12520401A, Naticchioni:2020kfb, DiGiovanni:2025oyo, Saccorotti:2023gnz} (Italy) or in the Meuse-Rhine (MR) region (Netherlands)~\cite{Bader:2022tdz, Koley:2022wpe} and a 2L shape configuration with the separated detectors in the two sites, with arm lengths from 15km to 20km and different orientations. Moreover, the possibility of adding cryogenic technology is under scrutiny, since it can add key information in the low frequency band. \\
    One of the greatest expectations for ET is the large number of detected events that will be the order of hundreds of thousands of compact binary coalescences per year, up to very high redshift (z$\sim$50)~\cite{Mancarella:2023infer}. These measurements will be pivotal because the large statistics will allow us to shed light on astrophysical population properties, and at the same time improve the cosmological applications of GW detections. However, such a large number of events will bring new challenges since it will require developing new, fast, and reliable data analysis tools. For the time being, in the spirit of contributing to the choice of optimal layout, the community has worked on fast tools to estimate the impact of different configurations on the science, developing tools based on Fisher Information Matrix (FIM) ~\cite{Vallisneri:2007ev}. Many analysis were already performed, e.g. ~\cite{Zhao:2017CBC, Pieroni:2022bbh, Iacovelli:2022For, Borhanian:2022czq}, with several codes publicly available: \texttt{GWFAST}~\cite{Iacovelli:2022GWF}, \texttt{GWBENCH}~\cite{Borhanian:2020GWB}, \texttt{GWFISH}~\cite{Dupletsa:2022GWF}, \texttt{GWDALI}~\cite{deSouza:2023DALI} and \texttt{TIDoFM}~\cite{Chan:2018TIdofm, Li:2021Tidofm}. For a comparison, see~\cite{Abac:2025BB}. 
    All these codes are \texttt{Python} based and employ different solutions to compute the FIM (or even going beyond the FIM approximation, e.g., \texttt{GWDALI}) on intrinsic and extrinsic parameters of GW events.
    In this paper, we present a new \texttt{Julia}-based tool to forecast the accuracy of CBC parameter estimation for different ET configurations. This new tool presents several advantages, such as the speed ($\sim 0.3$ sec per source for FIM computation) and the fact that it does not rely on external libraries, making the code self-consistent, easy to use and modify. Similarly to~\cite{Iacovelli:2022GWF}, the code makes use of \textit{automatic differentiation} thanks to state-of-the-art frequency-domain waveform models written entirely in \texttt{Julia}. The code supports several waveforms, e.g., \texttt{IMRPhenomXHM}~\cite{Garcia-Quiros:2020XHM}, \texttt{IMRPhenomXAS}~\cite{Pratten:2020XAS}, \texttt{IMRPhenomD}~\cite{Husa:2015PhD, Khan:2015PhD} and \texttt{IMRPhenomHM}~\cite{London:2017HM} for BBH (i.e., no tidal deformation allowed), \texttt{IMRPhenomD\_NRTidalv2}~\cite{Dietrich:2019NRTidal} for BNS and \texttt{IMRPhenomNSBH}~\cite{Pannarale:2015NSBH, Thompson:2020PhenomNSBH} for NSBH. 
    For different configurations, we forecast the signal–to–noise ratio (SNR) distribution for the different CBCs and the parameter reconstruction accuracy. 
    Moreover, we assess the capabilities of the different detector configurations by investigating the joint constraints on pairs of parameters that are key in the various science cases. For example, measuring the Hubble parameter using the dark sirens method requires accurate determination of both the sky position and luminosity distance~\cite{Schutz:1986gp, DelPozzo:2011vcw, LIGOScientific:2021aug}. Thus, we assess the fraction of events that satisfy specific precision thresholds simultaneously.
    For each detector network, we consider four combinations of parameters: chirp mass and symmetric mass ratio, chirp mass and luminosity distance, chirp mass and sky localization, and finally, luminosity distance and sky localization. In each case, we quantify the percentage of events that achieve a given level of accuracy in one parameter, in the other, and in both simultaneously. Then, we explore the interplay between detector geometry and parameter estimation, examining the spatial distribution of 
    the events across the sky for different ET configurations. We analyze the number of sources that meet two representative precision thresholds, and emphasize the importance of considering not only individual parameter performance, but also their spatial correlation across the sky when assessing detector configurations. \\
    Finally, to evaluate the validity and the performance of FIM forecasts for 3G detectors, we compare with a full Bayesian inference using a novel Hamiltonian Monte Carlo (HMC) sampler~\cite{Neal:2011HMC} informed by the FIM itself. As a first attempt, we perform the comparison under simplifying assumptions, such as zero noise injection and a high SNR, to isolate the performance of the sampler in a controlled setting. We use a network of three detectors, combining the 2L ET rotated configuration with a 40km CE located in the US. Even if at a native stage, we demonstrate that the FIM-informed HMC can produce high-fidelity samples at a fraction of the computational cost of traditional methods.

    The paper is structured as follows. In~\cref{sec:formalism}, we review the FIM formalism. In~\cref{sec:detector_catalog}, we introduce the detector configurations and the catalogs used. Later, in~\cref{sec:gwjulia}, we introduce the code \texttt{GWJulia} and, in~\cref{sec:forecast}, we perform the FIM analysis and show the results. Then, in~\cref{sec:MCMC}, we show the results of the HMC analysis in comparison with our FIM results. Finally, in~\cref{sec:conclusions}, we draw the conclusions of this work. 

\section{Formalism}\label{sec:formalism}
    The output of an interferometer is a datastream $d(t)$, which can be expressed as  
    \begin{equation}\label{eq:signal}
        d(t) = s(t) + n(t)\,,
    \end{equation}
    where $s(t)$ contains the signal and $n(t)$ represents the noise, which we will characterize in~\cref{sec:detector_catalog}. Concerning the signal, this is the result of the projection of the GW $h$ on the detector, where $h$ is 
     \begin{equation}
        h \equiv h_{i j} = h_+e_{ij}^{+} + h_\times e_{ij}^{\times}\,, 
    \end{equation}
    where $e_{ij}^{+/\times}$ are the polarization tensors and $h_{+/\times ij}$ are the two GW polarizations, e.g., see~\cite{maggiore2008gravitational}. 
    Then the projection is performed using the detector tensor $D_{ij}$, which results in
    \begin{equation}
        s = h_+e_{+ij} D^{ij}+ h_\times e_{\times ij} D^{ij} \equiv  h_+F_+ + h_\times F_\times\,,
    \end{equation}
    where $F_{+/\times\,ij}$ are the detector pattern functions \footnote{Notice that this expression assumes the long wavelength approximation to hold. In fact, the high frequencies reached by low mass events and the long arms of the third-generation interferometers can violate this approximation. This might be a sizable effect, in particular for the proposed 40km CE~\cite{Evans:2021CE} as highlighted by~\cite{Essick:2017wyl}, while it is negligible for the 15km arms of ET.}.
    If we neglect higher order harmonics for simplicity, the polarizations $h_{+/\times}$ can be expressed as
    \begin{equation}
        h_+ = A \cos\iota \,e^{i\Phi}\,,\quad h_\times = A \frac{1+\cos^2\iota}{2}\, e^{i\Phi}\,,
    \end{equation}
    where the amplitude $A$ is obtained directly from the waveform model, $\iota$ is the inclination angle and will be characterized in~\cref{sec:detector_catalog}, while the phase $\Phi$ is a combination of several terms. In particular, $\Phi$ includes a term from the waveform, terms related to the time to coalescence and phase of coalescence, and terms related to the Earth's motion during the inspiral phase. These expressions are modified in the presence of higher harmonics~\cite{Iacovelli:2022For}, and a sum over the higher terms needs to be included.\\
    In this work, we rely on the FIM formalism, which, offering a reasonable balance between efficiency and accuracy, is extensively applied to the GW analysis~\cite{Borhanian:2020GWB, Dupletsa:2022GWF, Iacovelli:2022GWF, Chan:2018TIdofm, Li:2021Tidofm, deSouza:2023DALI}. For a comparison of the different codes, see~\cite{Abac:2025BB}, while for a review of the literature on the topics and the limits of applicability of the FIM, see~\cite{Cutler:1994ys, Vallisneri:2007ev, Rodriguez:2013mla, Finn:1992wt}. 
    Before introducing the FIM, we introduce the Signal-to-Noise (SNR), which gives a measure of the strength of the signal and, for a single detector, is defined as
    \begin{equation}
        {\rm SNR} = (s\mid s)^{1/2}\,,
    \end{equation}
    where
    \begin{equation}\label{eq:scalar_prod}
        (a \mid b)=2 \int_{f_1}^{f_2} \frac{a(f) b^*(f)+a^*(f) b(f)}{S_n(f)} \mathrm{d} f,
    \end{equation}
    is a noise-weighted scalar product, with $S_n(f)$ representing the detector noise power spectral density (PSD) and the symbol $*$ is the complex conjugate. The specific noise levels for the different detectors considered in this work are reported in~\cref{sec:detector_catalog}. For a network of detectors, the SNR is
    \begin{equation}\label{eq:network_SNR}
        {\rm SNR}_{network} = \left[\sum_i {\rm SNR}_i^2\right]^{1/2}\,,
    \end{equation}
    where $i$ indicates the different detectors. In this work, we consider a signal to be detected when the network SNR is above a threshold, that we take to be $\rm SNR_{thres}$ = 12 as motivated in~\cite{Iacovelli:2022For}.\\
    Under the assumption that the noise is stationary, Gaussian distributed and with zero mean, we introduce the FIM defined as
      \begin{equation}\label{eq:FIM}
       \Gamma_{i j}\equiv -\left\langle \frac{\partial^2\log\mathcal{L}(d|\boldsymbol{\theta}) }{\partial \theta^i \partial \theta^j}\right\rangle_n=\left(\left.\frac{\partial s}{\partial \theta^i} \right\rvert\, \frac{\partial s}{\partial \theta^j}\right)\,,
    \end{equation}
    where $\theta^i$ represents one of the CBC parameters $\boldsymbol{\theta}$ and $\log\mathcal{L}$ is the standard Gaussian likelihood (see e.g.~\cite{Rodriguez:2013mla}) of the datastream $d$ measured at the detector given the event parameters $\boldsymbol{\theta}$. Moreover, $\langle\dots\rangle_n$ indicates the ensemble average over the noise realizations. 
    As carefully shown in \cite{Vallisneri:2007ev, Rodriguez:2013mla, Cornish:2006ry}, in the high-SNR limit, the covariance of the parameters under the posterior converges to the inverse of the FIM.
    However, this convergence comes with some caveats. In fact, GW likelihoods are usually multimodal and can show significant non-Gaussian tails, even with high SNRs. Moreover, some parameters, e.g., spins and tidal deformabilities, are more prone to non-Gaussian behaviours. In particular, the sensitivities to spins are usually low and the FIM can predict errors that go outside the physical bounds, i.e., $\chi\in[-1,1]$. A possible way to partially solve this issue could be the one described in \cite{Dupletsa:2024gfl}. Moreover, when dealing with tidal deformabilities, non-Gaussianities can arise due to the strong degeneracy between the tidal deformabilities of the two bodies. 
    In addition to these difficulties, inverting FIM can be problematic, particularly in the BNS case, since two additional parameters are involved~\cite{Vallisneri:2007ev}. Another limitation of the inversion is the case of face-on events, in which the GW from the event is nearly circularly polarized, i.e., the plus and cross polarizations have nearly the same amplitude. Thus, this can induce degeneracies among inclination, distance, phase-to-coalescence, and polarization angle; definitions are provided in \cref{sec:detector_catalog}. Another possible issue is that the inversion can be negatively affected even by a single poorly constrained parameter, i.e., the FIM would have a row-column of numbers close to zero, increasing numerical errors.
    One justification for using the FIM in our context is that it is expected to yield better results when analyzing quantities across a large number of events. For example, individual low-SNR or face-on events may be biased; however, the overall accuracy of a parameter inferred from a catalog of many events will be close to the true distribution obtained from a full parameter estimation. In \cite{Begnoni:2025mtz}, as few as ten sources were enough to recover the Bayesian results.

    \section{Detector and catalog} \label{sec:detector_catalog}
    In this section, we introduce and characterize the properties of the detectors with a focus on how to include the Earth's rotation. Then we will comment on the catalog properties and the waveforms used to generate the signals.
    
    \subsection{Detector properties}
\begin{figure}[t!]
        \centering
        \includegraphics[width=0.7\linewidth]{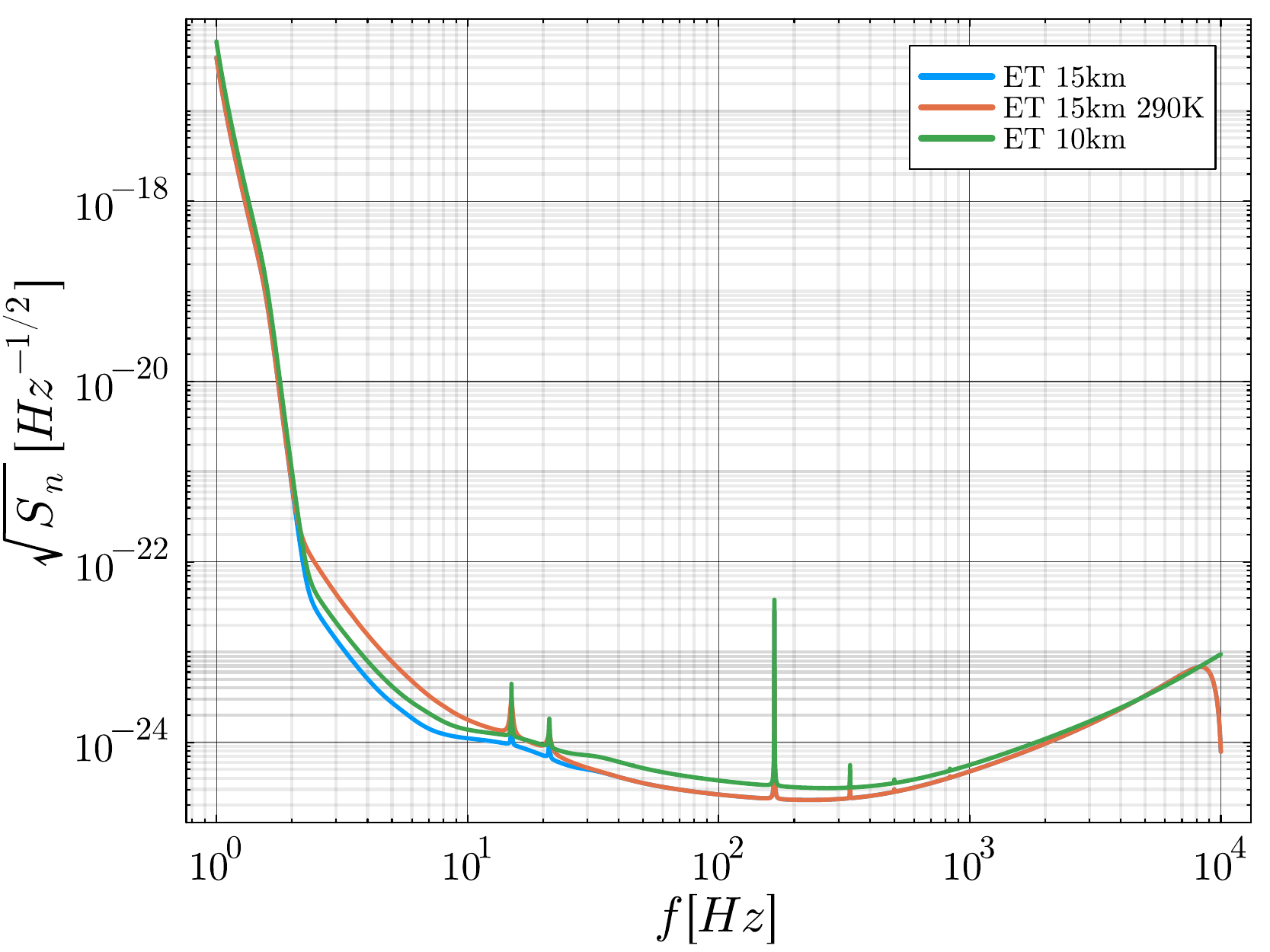}
        \caption{Plot of the detector amplitude spectral densities (ASD) for the three ET designs used in this work. In blue, there is the 15km arm length with cryogenic, in orange, the 15 km non-cryogenic, while in green, the 10 km cryogenic.}
        \label{fig:sensitivity-curves}
    \end{figure}

    Each detector is determined by its shape, position on Earth, orientation, and sensitivity curve. The shape, i.e., the geometry, is defined through the opening angle between the arms $\zeta$, which is $90^{\circ} $ for L-shaped and $60^{\circ} $ for triangular detectors. The latitude $\lambda$ and longitude $\varphi$ define the position of the center of the detector, i.e., the location of the beam-splitter for L-shaped and the center of the triangle for triangular detectors. Finally, the configuration is fully specified through the orientation angle, representing a rotation in the detector plane. 
    In this work, we define the orientation as the angle between the arm bisector of the detector and the local East and refer to it as $\gamma$. Following the convention used in~\cite{Iacovelli:2022For}, in the following, we collectively denote all these parameters with $\bm{\lambda}=\{\lambda,\varphi,\zeta,\gamma\}$ and express the pattern functions as
    \begin{equation}
    \begin{aligned}
    & F_{+}(t , \theta, \phi, \psi, \boldsymbol{\lambda})=\sin \zeta[a(t , \theta, \phi, \boldsymbol{\lambda}) \cos 2 \psi+b(t , \theta, \phi, \boldsymbol{\lambda}) \sin 2 \psi]\,, \\
    & F_{\times}(t , \theta, \phi, \psi, \boldsymbol{\lambda})=\sin \zeta[b(t , \theta, \phi, \boldsymbol{\lambda}) \cos 2 \psi-a(t , \theta, \phi, \boldsymbol{\lambda}) \sin 2 \psi]\,,
    \end{aligned}
    \end{equation}
    where $a(t , \theta, \phi, \boldsymbol{\lambda})$ and $b(t , \theta, \phi, \boldsymbol{\lambda})$ are two time-dependent functions derived in~\cite{Jaranowski:1998qm} and $\psi$ is the polarization angle, which will be characterized later in this section.
    The rotation of the Earth introduces other effects~\cite{Cutler:1997ta, Cornish:2003vj, Baral:2023xst, Chen:2024kdc} that need to be taken into account in the FIM evaluation.
  The inclusion of such effects can help break some degeneracies, particularly in the case of the long-lived signals, such as BNSs. On the other hand, not modelling these effects properly might introduce artifacts in the parameter estimation, see, e.g.,~\cite{Cutler:1997ta, Cornish:2003vj}. The procedure that we apply in our code to include these effects follows~\cite{Iacovelli:2022For}. We consider the two leading effects: an amplitude and phase modulation due to the pattern functions and a Doppler shift. The first effect originates from the time dependency of $F_{+/\times}(t)$, which introduces a modulation between the different frequencies (since they are measured by the detector at different times). While, the Doppler shift originates from the fact that different frequencies take different times to reach the detector, due to its motion. This effect is negligible for the amplitude, but is instead relevant for the phase and is effectively a frequency-dependent phase shift depending on the detector position on the Earth's surface and the frequency of the Earth's rotation. \\
    Since ET's final design and configuration have not been decided yet~\cite{Branchesi:2023COBA, Abac:2025BB}, in this paper, we consider different setups to compare their capabilities in detecting and estimating the parameters of CBCs.
    For this purpose, we consider, in~\cref{fig:sensitivity-curves}, three noise PSDs $S_n$:
    \begin{itemize}
        \item In green, we show the PSD for $L=10 \rm km$ and cryogenic technology. 
        \item In blue, we show the PSD for $L=15 \rm km$ and cryogenic technology.
        \item In orange, we show the PSD for $L=15 \rm km$ without cryogenic technology.  
    \end{itemize}
    The triangular detector is formed by three detectors with opening angle $\zeta = 60^{\circ}$.
    The current network proposals contemplate~\cite{Branchesi:2023COBA, Abac:2025BB} a triangular design or 2L networks. In the latter case, there is also the freedom of the orientation between the two detectors $\beta$, which affects the precision of the parameter estimations~\cite{maggiore2008gravitational}. The maximum sensitivity for CBC parameter estimation is achieved when both polarizations (i.e., plus and cross) are measured. The optimal configuration to achieve this outcome corresponds to two detectors misaligned with an angle of $\beta_{\rm CBC} = \pi/4 + n\pi/2$, where $n$ is an integer\footnote{Notice that this statement is only valid statistically, i.e., when dealing with a large number of sources. Indeed, some CBCs may be better measured with a different orientation angle between the two detectors. }. Moreover, a large separation (ideally at opposite sides of the Earth) between the detectors improves the sky localization of the sources. Let us add that for gravitational waves background (GWB), the most sensitive configuration corresponds to two aligned detectors, i.e., $\beta_{\rm GWB}=n\pi/2$~\cite{Branchesi:2023COBA}. Thus, determining the orientation choice requires considerations beyond only the CBC sources but should be based on the different science cases that the collaboration aims to address~\cite{Ebersold:2024ori, Abac:2025BB, Branchesi:2023COBA}.
    Each detector has an orientation $\gamma$ w.r.t. a given direction, which in this work (and in \texttt{GWJulia}) we choose w.r.t. the local East. In the coplanar limit, to maximize the sensitivity to CBC, one would like that $\beta = \beta_{\rm CBC} = \gamma_2-\gamma_1$, where $\gamma_i$ indicates the orientation of the $i$-th detector.
    When instead the detectors are not coplanar, as two detectors on Earth, due to Earth curvature, the difference of the orientations does not correspond to $\beta_{\rm CBC}$. For such a purpose, the orientations should be specified w.r.t. the great circle passing through the two detector sites~\cite{Flanagan:1993ori, Christensen:1996origwb}, which we identify with $\alpha$. Then to achieve the maximum sensitivity for CBC one needs that $\alpha = \alpha_{\rm CBC} = \gamma_2-\gamma_1$, where $\alpha_{\rm CBC}$ takes into account that the two detectors are not coplanar. This leads to $\alpha_{\rm CBC}= \beta_{\rm CBC}+2.51^{\circ}$ for two detectors, one in Sardinia, one in the MR region. 
    On the other hand, the optimal $\alpha$ for the GWB is $\alpha_{\rm GWB}=2.51^\circ$, which is also the statistically worst orientation for CBC. In this work, we will see that there are some caveats to this statement.
    The networks we analyze are: 
    \begin{itemize}
        \item \texttt{T}: triangular ET with 10 km arms in Sardinia featuring cryogenic technology
        \item \texttt{2L\_0}: two 15 km L-shape interferometers, one in Sardinia, one in the MR Euroregion, both with cryogenic technology. The orientations of the two detectors are chosen such that $\beta=0^\circ$.
        \item \texttt{2L\_45}: same as \texttt{2L\_0} with the exception that the orientations lead to $\beta=45^\circ$. This is expected to be the best network for measuring CBCs.
        \item \texttt{2L\_290K\_0}: two 15 km L-shape interferometers, one in Sardinia, one in the MHR Euroregion, and only the one in Sardinia features the cryogenic technology. The orientations are chosen such that $\beta = 0^\circ$.
        \item \texttt{2L\_290K\_45}: same as \texttt{2L\_290K\_0} with the exception that the orientations lead to $\beta=45^\circ$. 
    \end{itemize}
    In this work, and in \texttt{GWJulia}, we also consider that the beam splitters presented in the triangular configuration have slightly different positions.

\subsection{Catalog properties }
    A CBC is identified by numerous parameters, both intrinsic, i.e., characterizing the source properties, and extrinsic, i.e., illustrating the relation between the detector and the source, which we collectively denote with $\boldsymbol{\theta}$. Among the intrinsics, in this work, we consider the masses $m_{1/2}$ and the longitudinal spins $\chi_{1/2\,z}$. In the case of BNS and NSBH, a tidal deformability $\Lambda_{1/2}$ term for each NS is also considered.
    Within the extrinsics, we instead consider: the inclination angle $\iota$, i.e., the angle between the binary angular momentum and Earth position, the sky position, indicated with $(\theta,\phi)$, the polarization angle $\psi$ and the time and phase to coalescence, indicated respectively with  $t_{\rm coal}$ and $\Phi_{\rm coal}$.\\
    Each source is characterized by these parameters and each of them needs to be drawn by a distribution.
    Given the population type, i.e., BBH, BNS, NSBH, we draw realizations for all the CBC parameters. When possible, we rely on the most recent populations provided by the LVK collaboration~\cite{KAGRA:2021pop}. 
    Let us review the properties of the populations starting from the parameters that do not depend on the type of CBC.
    The sky positions are drawn from a uniform distribution over the sphere, i.e., $\varphi$, the azimuthal angle is drawn from a uniform distribution, and $\theta$, the polar angle, is uniform in the cosine\footnote{In reality, since we expect that astrophysical CBCs reside mainly in galaxies, sky positions $(\varphi,\theta)$ and the inclination angle $\iota$ should, at least in principle, be linked to the distribution and orientation of their main plane of galaxies. However, given the low measurement precision, which is insufficient to appreciate the underlying distribution, which should resemble the distribution of Large Scale Structures (LSS), we still consider these variables to be drawn from uniform distributions. }. Similarly to $\theta$, the inclination angle $\iota$ is drawn from a uniform distribution in the cosine. 
    Then, the polarization angle $\psi$ represents the rotation needed to align the source frame, where the $h_{+/\times}$ polarizations are defined, with the detector frame~\cite{whelan2013geometry}. This angle is sampled uniformly in $[0,\pi]$.
    The time to coalescence, $t_{\rm coal}$, represents the time of the merger and is uniform in a day. The phase to coalescence $\Phi_{\rm coal}$ represents the phase at which the merger happens and, similar to $t_{\rm coal}$, it is drawn from a uniform distribution, this time in the interval $[0,2\pi]$.
    Finally, for the redshift distribution, we stick to~\cite{Madau:2014Dickinson}, which is convoluted with a time delay distribution. The time delay represents the time between the star formation and the merger of the CBC, and we consider a power-law distribution. A key and still uncertain parameter in this distribution is the minimum time delay allowed $t_{d,\rm{min}}$, which we consider 20 Myrs for all types of sources~\cite{Mapelli:2017hqk, Regimbau:2012ir, Belgacem:2019tbw}. \\
    Let us proceed by discussing the distribution of the parameters having different pdfs depending on the CBC type. Considering first the BBH, the masses are drawn from a \texttt{PowerLaw + Peak} distribution~\cite{KAGRA:2021pop}, which consists of a power law plus a gaussian peak at $M_{\rm peak}\approx 28 M_{\odot}$. 
    On the other hand, BNSs are drawn a uniform distribution~\cite{KAGRA:2021pop} for the masses. The limits of this distribution $M_{\rm min}^{\rm BNS} = 1 \,M_{\odot}$ and $M_{\rm max}^{\rm BNS}= 2.15 \, M_{\odot}$ are still uncertain~\cite{Suwa:2018uni, Kacanja:2024hme, LIGOScientific:2024NSBH}. The higher limit is chosen to accommodate the equation of state (EoS) we chose for our analysis, which we will define better later in this section.
    While for the case of NSBHs the BH mass is drawn according to~\cite{Giacobbo:2018NSBH, Zhu:2020NSBH} and the NS mass is, instead, drawn from a normal distribution $\mathcal{N}(\mu=1.33, \sigma=0.09)$ as described in~\cite{Iacovelli:2022For, Broekgaarden:2021NSBH}.
    Regarding the spins, in the case of BBH, we employ the Default model from~\cite{KAGRA:2021pop}. Instead, for BNS and the NS component of the NSBH, we draw it uniformly in the interval [-0.05,0.05] due to the low spins expected~\cite{Miller:2014aaa, Pacilio:2021jmq}. While for the BH component of the NSBH, we draw it from a normal distribution $\mathcal{N}(\mu=0,\sigma=0.15)$~\cite{Iacovelli:2022For, Broekgaarden:2021NSBH}.
    A recap of the distributions for all the parameters is reported in~\cref{tab:CBC_parameters}. \\
    The NSs have an additional parameter w.r.t. the BHs, the dimensionless Love number~\cite{Hinderer:2007mb}, also known as tidal deformability. 
    This additional parameter produces tidal corrections which are implemented in the waveform models \texttt{IMRPhenomD\_NRTidalv2} and \texttt{IMRPhenomNSBH}, both in the phase and amplitude, see~\cref{app:waveform}.
    Each CBC is associated with a dimensionless tidal deformability $\Lambda_i$ for each component of the binary. The tidal deformability is a function of the mass and radius of the star, which is in turn a function of the EoS of the star. It gives a measure of the deformability of the body during the coalescence when placed in a non-uniform gravitational field. For BHs, it is fixed to zero, while for NS, depending on the EoS, the tidal deformability is of order $\mathcal{O}(10^3)$ at typical NS masses. In a more realistic setup~\cite{Zhao:2018nyf}, which goes beyond the scope of the current analysis, this quantity should also depend on the spin of the star. Once an EoS is chosen, the tidal deformability is a function of the mass only. Our choice for the NSs in BNS is the APR4~\cite{Akmal:1998cf, read2009constraints, Ozel:2016oaf} EoS, which is allowed by GW170817~\cite{LIGOScientific:2017_0817} and can support NS masses up to our limit of $M_{\rm NS}^{\rm max}=2.15 \, M_\odot$. Moreover, even though the difference between various EoS is significant~\cite{Pacilio:2021jmq}, their effect on the other parameter estimations is limited, since the BNS waveform is calibrated against different EoS.\\
    For what concerns the NS in NSBH, to favor an easier comparison with \cite{Iacovelli:2022For}, we sample the tidal deformability from a uniform distribution Uniform(0, 2000).
    During the inspiral, the waveforms depend on the tidal deformability through the parameters $\tilde{\Lambda}$ and $\delta\tilde{\Lambda}$
\begin{align}\label{eq:lambdatilde}
    \begin{aligned} \tilde{\Lambda} & =\frac{8}{13}\left[\left(1+7 \eta-31 \eta^2\right)\left(\Lambda_1+\Lambda_2\right)+\sqrt{1-4 \eta}\left(1+9 \eta-11 \eta^2\right)\left(\Lambda_1-\Lambda_2\right)\right] \\ 
    \delta \tilde{\Lambda} & =\frac{1}{2}\biggl[\sqrt{1-4 \eta}\left(1-\frac{13272}{1319} \eta+\frac{8944}{1319} \eta^2\right)\left(\Lambda_1+\Lambda_2\right)\\   
    & \quad\quad +\left(1-\frac{15910}{1319} \eta+\frac{32850}{1319} \eta^2+\frac{3380}{1319} \eta^3\right)\left(\Lambda_1-\Lambda_2\right)\biggl]\end{aligned}\,,
    \end{align}
    which enter at 5-Post-Newtonian (PN) and 6PN order respectively. These corrections are implemented directly only in the phase of \texttt{TaylorF2} while \texttt{IMRPhenomD\_NRTidalv2} and \texttt{IMRPhenomNSBH} use a more complicated implementation calibrated on numerical relativity (NR) simulations, see~\cref{app:waveform}. However, even when calibrated to NR simulations, it is still valid that the 5PN correction $\tilde{\Lambda}$ is the leading term, while $\delta \tilde{\Lambda}$ is the next-to-leading order correction. The FIM is calculated by \texttt{GWJulia} over $\Lambda_{1/2}$, after that, the matrix is inverted and~\cref{eq:lambdatilde} is applied. 
    \\
    We also emphasize that the total number of events depends sensitively on parameters that are still poorly constrained. For example, the local merger rate, the distributions of time delays and the redshift evolution are not well constrained by current observations \cite{KAGRA:2021pop, Fishbach:2021mhp, Schiebelbein-Zwack:2024roj, Fishbach:2023pqs}. Moreover, their modelling remains affected by significant uncertainties \cite{Mapelli:2021gyv, Bouffanais:2020qds}. For the time being, relying on the current knowledge, even under pessimistic assumptions, ET is expected to detect $\mathcal{O}(10^4)$ sources per year \cite{Babak:2023lro} and thus not significantly affect the conclusions of our investigation. All the total number of sources remain consistents with previous analysis~\cite{Bellomo:2021mer, Iacovelli:2022For, Pieroni:2022bbh, Muttoni:2023prw, Borhanian:2022czq} given the LVK bounds~\cite{KAGRA:2021TC3}.
    In this work, we do not consider the overlap of different signals, which would complicate the parameter estimation. Furthermore, no effects of the revolution around the Sun are included, since they are subdominant~\cite{Iacovelli:2022For}. \\
    \begin{table}[h!]
        \centering
    \begin{tabular}{|c|c|c|c|}
    \hline Parameter & $\mathrm{BBH}$ & BNS & $\mathrm{NSBH}$ \\
    \hline$m_1$ & \multirow{2}{*}{ Power Law + Peak~\cite{KAGRA:2021pop}} & \multirow{2}{*}{ uniform in $[1,2.15] \mathrm{M}_{\odot}$} & fit reported in~\cite{Zhu:2020NSBH} \\
    \cline{1-1}\cline{4-4}$m_2$ & & & $\mathcal{N}(1.33,0.09)$ in $\mathrm{M}_{\odot}$\\
    \hline$z$ & \multicolumn{3}{|c|}{ Madau-Dickinson $+P\left(t_d\right) \propto 1 / t_d$} \\
    \hline $t_{d, \min }$ & 20 Myr & 20 Myr & 20 Myr \\
    \hline$d_L$ & \multicolumn{3}{|c|}{ computed from $z$ in Planck18 flat $\Lambda$CDM~\cite{Planck:2018vyg} } \\
    \hline$\chi_{1, z}$ & \multirow{2}{*}{ DEFAULT~\cite{KAGRA:2021pop}} & \multirow{2}{*}{ uniform in $[-0.05,0.05]$} & $\mathcal{N}(0,0.15)$ \\
    \cline{1-1}\cline{4-4}$\chi_{2, z}$ & & & uniform in $[-0.05,0.05]$ \\
    \hline$\chi_x, \chi_y$ & \multicolumn{3}{|c|}{0}   \\
    \hline$\Lambda_1$ & 0&\multirow{2}{*}{ APR4~\cite{Akmal:1998cf}} & 0 \\
    \cline{1-2}\cline{4-4}$\Lambda_2$ &0 & & Uniform(0,2000)\\
    \hline$\theta$ & \multicolumn{3}{|c|}{$\cos (\theta)$ uniform in $[-1,1]$} \\
    \hline$\phi$ & \multicolumn{3}{|c|}{ uniform in $[0,2 \pi]$} \\
    \hline$\iota$ & \multicolumn{3}{|c|}{$\cos (\iota)$ uniform in $[-1,1]$} \\
    \hline$\psi$ & \multicolumn{3}{|c|}{ uniform in $[0, \pi]$} \\
    \hline$t_c$ & \multicolumn{3}{|c|}{ uniform in 1 day (no overlap, no Earth revolution)} \\
    \hline$\Phi_c$ & \multicolumn{3}{|c|}{ uniform in $[0,2 \pi]$} \\
    \hline
    R(z=0) & 17 [Gpc$^{-3}$yr$^{-1}$] & 105.5 [Gpc$^{-3}$yr$^{-1}$] & 45 [Gpc$^{-3}$yr$^{-1}$] \\
    \hline
    $n_{\rm sources}$ $yr^{-1}$ &  $3.35 \times 10^4$ & $2.22 \times 10^5$ & $8.89 \times 10^4$ \\
    \hline
    \end{tabular}
        
        \caption{ Recap of all the values entering the catalog definition. These are the default settings of the catalog generation in \texttt{GWJulia}}\label{tab:CBC_parameters}
    \end{table}

\subsection{Waveforms}

    In the literature, there are many waveform models with different purposes, levels of accuracy, and calibration ranges. Here we present briefly the waveforms used in this work, while in~\cref{app:waveform} we give a more thorough explanation of their features. 
    All the waveforms used are part of the \texttt{IMRPhenom} category. These waveforms are computed in Fourier space and under the stationary phase approximation during the inspiral, i.e., the phase changes faster than the variations of the amplitude. \texttt{IMR} stands for inspiral-merger-ringdown, implying these are full phenomenological waveforms tuned with numerical relativity simulations.
    \begin{itemize}
        \item BBH: $\texttt{IMRPhenomXHM}$~\cite{Garcia-Quiros:2020XHM}, this waveform contains higher-order harmonics and mode mixing, making it the most advanced waveform we have at our disposal.
        \item BNS: $\texttt{IMRPhenomD\_NRTidalv2}$~\cite{Dietrich:2019NRTidal}, is the tidal extension of $\texttt{IMRPhenomD}$, allowing for the Love number $\Lambda$ to be different from 0.
        \item NSBH: $\texttt{IMRPhenomNSBH}$~\cite{Pannarale:2015NSBH, Thompson:2020PhenomNSBH}, builds on BBH waveforms with the possibility of having different types of mergers due to the behaviour of the NS (e.g., with or without the formation of a torus).
    \end{itemize}
    All of these waveforms represent state-of-the-art models~\cite{LIGOScientific:2024NSBH, KAGRA:2021TC3}. 
    Although the BBH waveform model adopted here does not include spin-precession effects, this is standard practice in the context of forecasts for future gravitational-wave detectors \cite{Branchesi:2023COBA, Abac:2025BB}. LVK events so far indicate that precession angles are generally small \cite{KAGRA:2021TC3}, and, while in principle they should be taken into account, the required implementation is technically demanding. Moreover, taking them into account would further ill-condition the Fisher matrix \cite{Vallisneri:2007ev, Berti:2004bd}. In fact, despite the physical potential of precession to break the spin mass ratio degeneracy \cite{Baird:2012cu, Chatziioannou:2014bma}, the local, linear nature of the FIM often fails to capture the resulting non-Gaussian posterior structures.
    Therefore, even if we are aware that this assumption could affect the other parameter precision, we will work with this simplifying approximation.

\section{GWJulia}\label{sec:gwjulia}

    \texttt{GWJulia} \github{https://github.com/andrea-begnoni/GW.jl} 
    is an open source code entirely written in \texttt{Julia} and based on automatic differentiation (AD) for derivatives. The code is completely self-contained (i.e., it does not rely on external libraries to perform heavy calculations), making it easier to understand and modify. The language \texttt{Julia} is compiled (like C) and designed for scientific computing. All these features lead to a very fast code with no loss of accuracy w.r.t. other codes available in the literature based on AD, e.g., \texttt{GWFAST}~\cite{Iacovelli:2022For, Iacovelli:2022GWF} and marginal loss of accuracy w.r.t. other frequency-domain codes based on finite difference methods like \texttt{GWBENCH}~\cite{Borhanian:2020GWB} and \texttt{GWFISH}~\cite{Dupletsa:2022GWF}. For a comparison, see~\cite{Abac:2025BB}.
    As an example, to calculate the FIM for an event with the most demanding waveform implemented yet (i.e., \texttt{IMRPhenomXHM}, which includes higher order harmonics), the code takes less than a second for a network of three interferometers. This allows us to evaluate the catalog of 1 year of observations, corresponding to $\mathcal{O}(4\cdot10^4)$ events, in less than two hours (considering 8 CPUs), making it feasible to analyze 3G detector catalogs on a laptop.
    The code relies on the \texttt{Julia} \texttt{ForwardDiff} package~\cite{Revels:2016for} to perform derivatives using AD. In a nutshell, \textit{automatic differentiation} is based on the idea that all operations are a sequence of basic arithmetic operations, which are differentiable (e.g., the derivative of a sum of two variables is the sum of the derivatives of those variables). This fact, together with a smart implementation of the chain rule, allows bypassing the usual issues with numerical derivatives and allows evaluating derivatives at machine precision.  
    While the GW waveform is a very complex function, it can still be decomposed into many arithmetical steps, leading to an ideal application of \textit{automatic differentiation} as already tested in~\cite{Iacovelli:2022GWF, Edwards:2023rip}.
    The use of \textit{automatic differentiation} is possible thanks to the use of state-of-the-art frequency-domain waveform models written entirely in \texttt{Julia}. The code supports all the waveforms presented in the previous section, i.e.,
     \texttt{IMRPhenomXHM}, \texttt{IMRPhenomXAS}, \texttt{IMRPhenomD}, and \texttt{IMRPhenomHM} for BBH (i.e., no tidal deformation allowed), \texttt{IMRPhenomD\_NRTidalv2} for BNS, and \texttt{IMRPhenomNSBH} for NSBH. 
    In its current version, the code provides a full pipeline from catalog generation to parameter estimation as follows:
    \begin{itemize}
        \item \textbf{Catalog definition}: the code generates a catalog of events according to the latest populations provided by the LVK collaboration~\cite{KAGRA:2021pop}. It can also read a catalog from a file since the inputs of the code are arrays of parameters (i.e., no particular structure is required)
        \item \textbf{Waveform generation}: the second step is to generate the waveform. It depends on the intrinsic parameters of the binary $\bm{\theta}_{\rm intr}=\{\Mc,\eta,\Vec{\chi}_1,\Vec{\chi}_2,\Lambda_1,\Lambda_2\}$. With the addition of the inclination angle $\iota$, if higher modes are considered.
        \item \textbf{Projection onto the detector}: the third step consists of the actual measurement of the signal by a single detector or network of detectors and involves the projection of the waveform on each detector, keeping Earth rotation into account. This step depends then on the extrinsic parameters of the binary $\bm{\theta}_{extr}=\{\theta,\phi,\iota,\psi,t_{\rm coal}, \Phi_{\rm coal}\}$ and on the detector configuration (i.e., $\bm{\lambda}=\{\lambda,\varphi,\zeta,\gamma\}$).
        \item \textbf{SNR evaluation}: then the code evaluates the SNR using only the waveform amplitude.
        \item \textbf{FIM calculation}: once the code verified that the SNR is above a certain threshold (${\rm SNR}>{\rm SNR}_{\rm thres}$), it performs the derivatives of the waveform needed for the FIM and then computes the integrals of the scalar product of~\cref{eq:scalar_prod}, for $a=b=s$.
        \item \textbf{FIM inversion}: the final step is the inversion of the matrix to obtain the covariance matrix. For this purpose, we use the Cholesky decomposition. This step can be challenging since the FIM can be close to singular (i.e., the determinant close to zero). 
    \end{itemize}
    An in-depth tutorial guiding the user through these steps is present at the GitHub link \github{https://github.com/andrea-begnoni/GW.jl}. 
    At any point in the pipeline, results can be exported to continue the analysis using other programming languages (e.g., \texttt{Python}) since there are no compatibility issues. The code also allows for easy addition of new waveform parameters, e.g., linked to beyond GR theories, making the FIM analysis straightforward.

\section{Forecasts on parameter estimation with different ET configurations} \label{sec:forecast}


    For each event type, we start by presenting the SNR results, followed by the FIM analysis. While typically, larger SNR translates to better parameter estimation, as we will see later, this is not always the case and is valid only on average. For example, two events with the same SNR but measured with two different networks will have very different parameter accuracies, e.g., the network with more detectors will likely have a better sky position accuracy.
    Overall, this introduces non-trivial behaviours in the FIM evaluation, which limits the possibility of establishing a clear relation between SNR and parameter accuracy. 
    
    However, it is still interesting to compare the SNR distribution for different detectors. Notice that while the \texttt{T} configuration has three interferometers, the other networks are composed of only two interferometers. 
    After the SNR, we will present the FIM analysis showing the cumulative distribution functions (CDF) relative to the parameter uncertainties, given the assumed population of CBC. In the case of BBH, we will further analyze the golden events detection implications.

\subsection{Binary Black Holes}

\begin{figure}
    \centering
    \includegraphics[width=\linewidth]{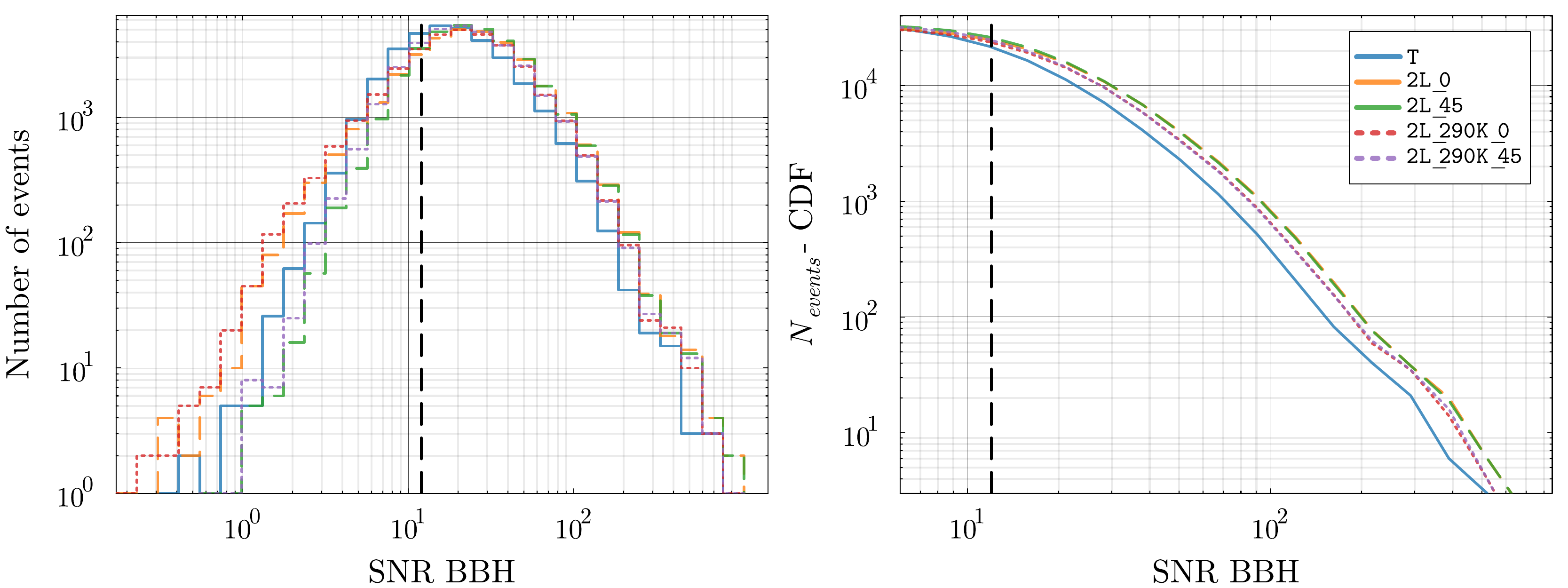}
    \caption{Left: SNR histogram for the BBH events. Right: CDF of the SNR for the same events. The vertical line represents ${\rm SNR_{thres}=12}$. The \texttt{T} detector in blue has the lowest SNR for detected events, while the \texttt{2L\_45} network in green dashed has the highest. The \texttt{2L\_0} and \texttt{2L\_290K\_0} networks, respectively in orange dashed and red dotted, have a tail of low SNR events due to their larger blind spots, compared to the other networks. The \texttt{2L\_290K\_45} network in purple dotted follows the \texttt{2L\_290K\_0} at high SNRs while it follows \texttt{2L\_45} at low SNRs.}
    \label{fig:BBH_SNR}
\end{figure}

    In~\cref{tab:BBH_SNR} and~\cref{fig:BBH_SNR}, we present the results for the SNRs of the five configurations. We can see that the \texttt{2L\_45} network outperforms the other networks in almost all intervals, while reaching a similar result to the network \texttt{2L\_0} for SNR$>50$. 
    This can be explained by the fact that the \texttt{2L\_45} network is optimally placed to measure most of the CBCs, due to the better sky coverage guaranteed by the two detectors being at a 45° orientation. 
    To see this, one can look at the role of the pattern functions in the SNR expression, where for the SNR of one detector one can write
    \begin{equation}
         {\rm SNR}   \propto \sum_{+/\times} F^2_{i, *}\,,
    \end{equation}
    where $F_{*}$ indicates $F_{+/\times}$. This combined with \cref{eq:network_SNR} leads to \texttt{2L\_45} having no blind spots, because of how the pattern functions combine when they are rotated by $45^\circ$. While \texttt{2L\_0} has some blind spots, which are captured by the difference in the number of detected sources. There is instead no difference when we look at the high SNR sources, due to the fact that they are represented by sources perpendicular to the detector plane. In fact, for these off-plane sources, the network orientation plays no role in the SNR evaluation.
    This will instead be relevant in the FIM analysis, where, as we will see in~\cref{subsec:golden}, the picture is more complex. 
    The networks without the cryogenic technology, i.e., \texttt{2L\_290K\_0} and \texttt{2L\_290K\_45}, are slightly worse than their cryogenic counterpart. This is because the cryogenic technology allows for a better sensitivity in the low frequency band, which leads to a larger SNR. The difference is of a few percentage points for SNR $>12$, but it increases for larger SNRs, reaching $\approx$ 20\% more sources detected with SNR $>100$.

    \begin{table}[h!]
        \centering
        \begin{tabular}{|c|c|c|c|c|c|}
        \hline
        Network & SNR $>$ 8 & SNR $>$ 12 & SNR $>$ 20 & SNR $>$ 50 & SNR $>$ 100 \\
        \hline
        \texttt{T} & 87.6 \% & 71.1 \% & 43.3 \% & 9.1 \% & 1.7 \% \\
        \texttt{2L\_0} & 89.3 \% & 78.6 \% & 56.5 \% & 15.7 \% & 3.6 \% \\
        \texttt{2L\_45} & 94.1 \% & 82.9 \% & 58.1 \% & 15.6 \% & 3.5 \% \\
        \texttt{2L\_290K\_0} & 87.4 \% & 75.5 \% & 52.3 \% & 13.4 \% & 2.9 \% \\
        \texttt{2L\_290K\_45} & 92.3 \% & 79.6 \% & 53.5 \% & 13.3 \% & 2.8 \% \\
        \hline
        \end{tabular}
        \caption{Table representing the percentage of events over a given SNR threshold, which is indicated in each column. Performance is similar across the different networks, while \texttt{2L\_45} obtains the best results in almost all intervals.}
        \label{tab:BBH_SNR}
    \end{table}

\subsubsection{Fisher Information Matrix analysis}
   
\begin{figure}
    \centering
    \includegraphics[width=\linewidth]{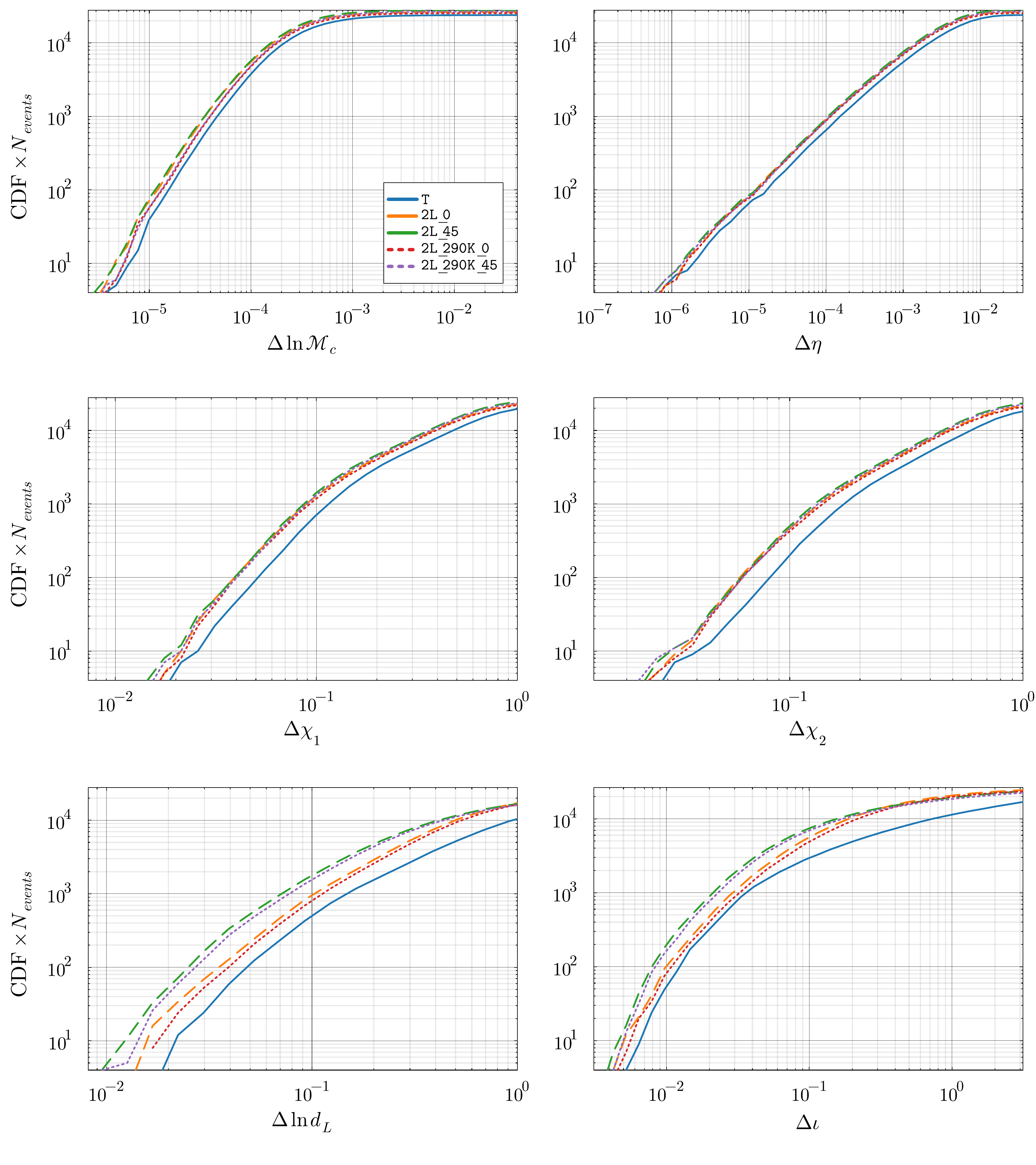}
    \caption{Forecast uncertainties for BBH, obtained with $\texttt{IMRPhenomXHM}$. The triangular configuration is represented in blue, the \texttt{2L\_0} in orange dashed, the \texttt{2L\_45} in green dashed, the \texttt{2L\_290K\_0} in red dotted, and the \texttt{2L\_290K\_45} in purple dotted. \textit{Top-left}: the CDF of the forecast relative uncertainty for the chirp mass $\mathcal{M}_c$.
    \textit{Top-right}: the CDF of the symmetric mass ratio $\eta$. \textit{Middle-left}: the CDF of the first spin $\chi_1$. \textit{Middle-right}: the CDF of the second spin $\chi_2$.
    \textit{Bottom-left}: the CDF of the relative uncertainty for the luminosity distance $d_L$. \textit{Bottom-right}: the CDF of the inclination angle $\iota$. The triangular configuration underperforms the others by roughly a factor of 1.5–2, whereas the \texttt{2L\_45} setup is marginally the most sensitive. Notably, \texttt{2L\_290K\_45} also outperforms \texttt{2L\_0} across a wide range of uncertainty levels.}
    \label{fig:BBH_CDF_parameters}
\end{figure}

\begin{figure}[h!]
    \centering
    \includegraphics[width=\linewidth]{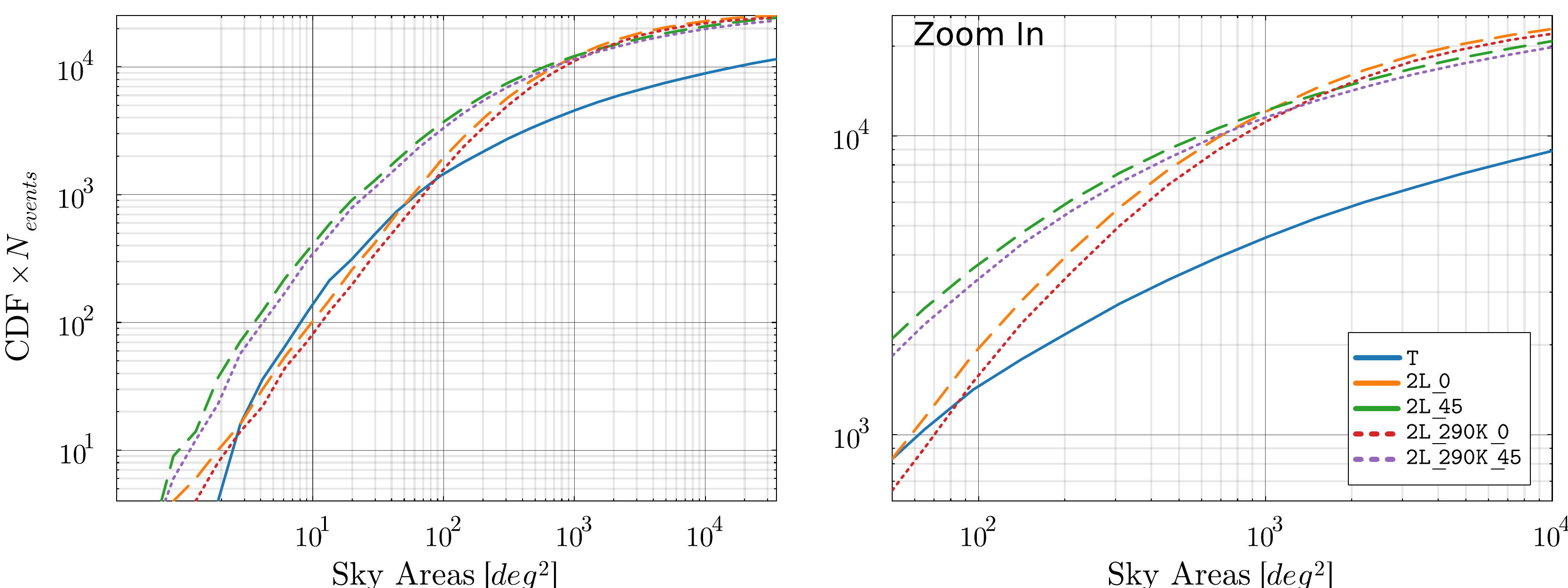}
    
    \caption{Left: CDF of the 90\% sky area for the BBH events. Right: a zoom-in of the left plot. An interesting behaviour emerges: the 45° networks outperform the other networks, but only for angular precision smaller than $10^3 \rm deg^2$, which corresponds to $\approx 30\%$ of the events. For better angular precisions, the \texttt{2L\_0} is the best network, suggesting that the choice of the best network is dependent on the scientific objectives.}
    \label{fig:BBH_sky}
    \end{figure}

    We proceed by considering the parameter uncertainties, forecasted with the FIM analysis. In~\cref{fig:BBH_CDF_parameters} we plot the CDF (multiplied by the number of events) of the forecast uncertainties for the relevant parameters. One of the best measured parameters is the chirp mass at the detector (i.e., the redshifted one), which is defined as
    \begin{equation}
        \mathcal{M}_{\rm c,det} = (1+z)\mathcal{M}_{\rm c,source} \,, \qquad\text{where}\qquad \mathcal{M}_{\rm c,source} = \frac{m_1^{3/5}m_2^{3/5}}{(m_1+m_2)^{1/5}} \,.
    \end{equation}
    While the symmetric mass ratio is defined as
    \begin{equation}
        \eta = \frac{m_1m_2}{(m_1+m_2)^2}\,.
    \end{equation}
    We notice that the best networks are \texttt{2L\_0} and \texttt{2L\_45}, followed by their 290K version and, finally, by the triangular design. 
    Similar behaviours are observed for the other parameters, with the other intrinsic parameters, i.e., the symmetric mass ratio $\eta$ and the spins, being similar across all 2L networks. The triangular network is worse by at most a factor of 2. 
    A different behaviour appears in the extrinsic parameters, e.g., the case of the luminosity distance $d_L$. Here, in fact, the 45° networks are better than their aligned counterpart. In particular, we notice that for most uncertainty levels, the \texttt{2L\_45\_290K} network performs better than the \texttt{2L\_0}. Interestingly, above a certain precision, the situation changes and the \texttt{2L\_0} network is better than the \texttt{2L\_45}. Moreover, this result agrees with~\cite{Branchesi:2023COBA}.
    In~\cref{fig:BBH_sky} and~\cref{tab:BBH_OmegaSky}, we notice a similar behaviour for the 90\% sky area, where the \texttt{2L\_45} networks outperform the other networks for most levels of uncertainty, while the \texttt{2L\_0} networks are better for uncertainties larger than $10^3 \,\rm deg^2$ which corresponds to $\approx 30\%$ of the total events. There are multiple competing effects determining this complex behaviour. The \texttt{2L\_45} networks are better at breaking the degeneracies, in particular the luminosity distance-incination one, given the fact that the pattern functions of the two detectors are rotated one with respect to the other by 45°. However, for events with lower SNR, this effect is compensated by the anisotropic pattern function of the \texttt{2L\_0} network. In fact, sky localization is affected by the derivative of the pattern functions, making an anisotropic sky coverage better at measuring this specific type of low-SNR events.

    This behaviour of the extrinsic parameters is important and poses new challenges. In fact, the decision on which network to choose should consider the scientific case one is examining, whether it is preferable to have many events with good precision or a few events with extremely good precision. We are going to discuss this in more detail in~\cref{subsec:golden}.

    \begin{table}[h!]
        \centering
    \begin{tabular}{|c|c|c|c|c|}
    \hline
    Network &  $\Omega$ $<$ 10 deg$^2$ & $\Omega$ $<$ 100 deg$^2$ & $\Omega$ $<$ 1000 deg$^2$ & $\Omega$ $<$ Whole sky  \\
    \hline
    \texttt{T} & 0.31 \% & 3.86 \% & 12.65 \% & 34.24 \% \\
    \texttt{2L\_0} & 0.23 \% & 4.68 \% & 32.91 \% & 75.43 \% \\
    \texttt{2L\_45} & 1.01 \% & 9.63 \% & 34.04 \% & 72.22 \% \\
    \texttt{2L\_290K\_0} & 0.18 \% & 3.79 \% & 30.14 \% & 72.49 \% \\
    \texttt{2L\_290K\_45} & 0.81 \% & 8.67 \% & 32.18 \% & 68.72 \% \\
    \hline
    \end{tabular}
        \caption{Table representing the percentage of events for which the 90\% sky areas are less than the threshold indicated in each column. 100\% represents the whole sources in the catalog (i.e., detected and not detected events). The 45° networks outperform the other networks, while the \texttt{T} design is comparable when considering the few high-precision sources. However, its performance degrades significantly for sources with precision worse than 1000 $\rm deg^2$.}
        \label{tab:BBH_OmegaSky}
    \end{table}


\subsubsection{Golden events}\label{subsec:golden}
\begin{figure}
    \centering
    \includegraphics[width=\linewidth]{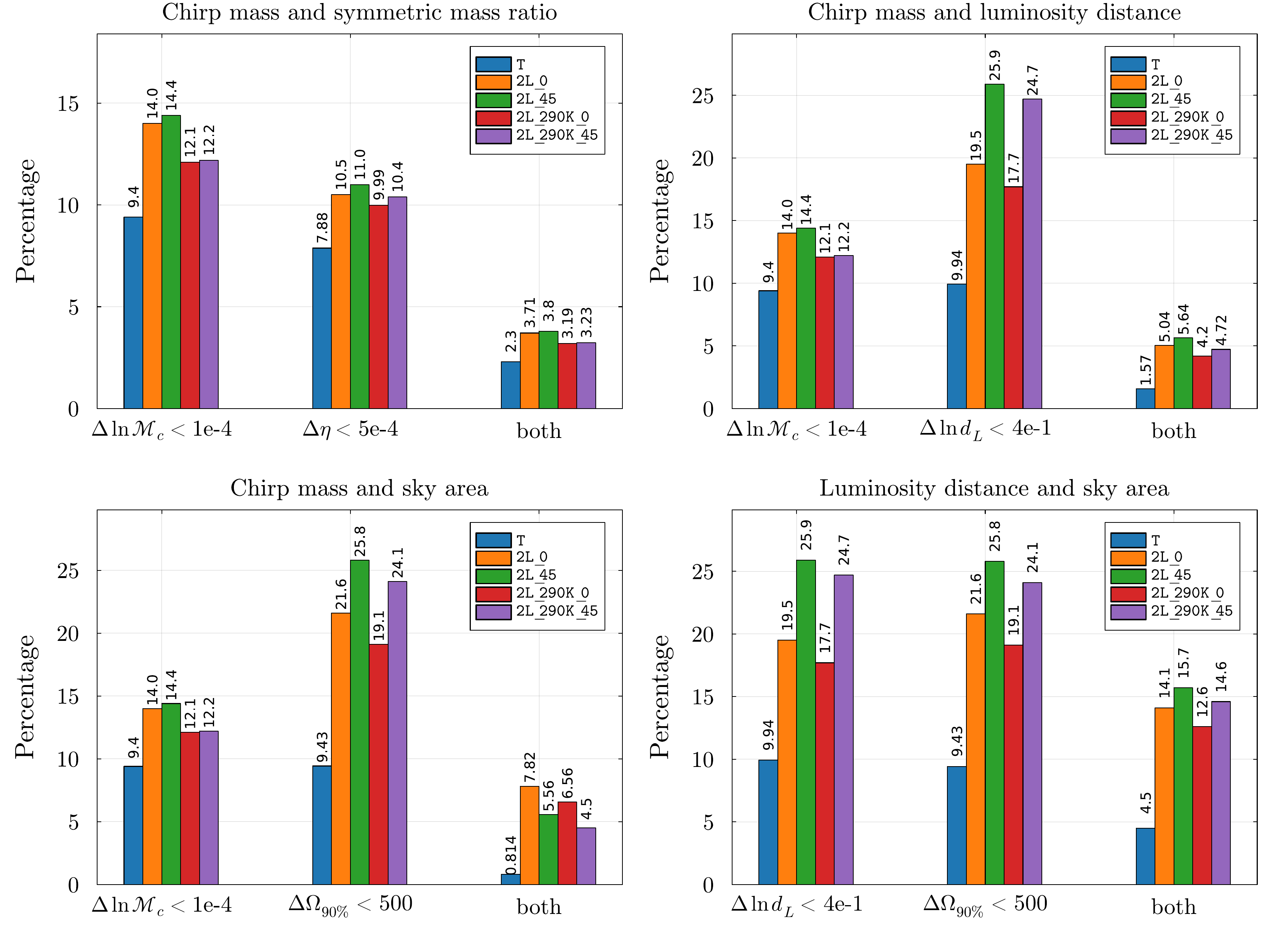}
    
    \caption{
    In each plot, there are three groups of data, containing 5 bars each (one bar per detector). Each group is identified by a condition. The y-axis shows the percentage of events with a precision better than the condition, e.g., the first group of the first plot represents the percentage of the total events with $\Delta\ln\mathcal{M}_c<10^{-4}$, the second group corresponds to the percentage of events with $\Delta\eta < 5\times 10^{-4}$ and the third group corresponds to the events which satisfy both conditions.
    \textit{Top-left:} Chirp mass and mass ratio. We can see that roughly only a third of the events satisfy both conditions. The hierarchy is maintained in each parameter, with the \texttt{2L\_45} being the best network and the triangular configuration being the worst.
    \textit{Top-right:} Chirp mass and luminosity distance. The \texttt{2L\_45} and the \texttt{2L\_290K\_45} networks are the best ones considering the luminosity distance; however,
    almost all of their advantage does not translate to a better measure of both parameters. 
    \textit{Bottom-left:} Chirp mass and sky precision. The \texttt{2L\_45} is the best network in each parameter, however, the \texttt{2L\_0} is better at measuring both parameters at the same time.
    \textit{Bottom-right:} Luminosity distance and sky precision. The \texttt{2L\_45} is the best network in each parameter, however the \texttt{2L\_0} is still comparable with \texttt{2L\_45} for the measurement of both parameters.}
    \label{fig:BBH_golden_events}
\end{figure}

We proceed by studying how accurately we can measure combinations of parameters. For this purpose, we set thresholds on the required accuracy for each parameter and check the percentage of events that pass these requirements. Naively, based solely on SNR consideration, one would expect the top 10\% of events, when ranked by any intrinsic parameter, such as the chirp mass, to largely coincide with the top 10\% ranked by another parameter, like the symmetric mass ratio.
However, this is usually not the case, and this behaviour is quite network dependent. We can see this in~\cref{fig:BBH_golden_events}, where we show the percentage of events with a precision better than a certain threshold for four different pairs of parameters. In particular, we consider: ($\Delta \ln \mathcal{M}_c,\Delta \eta)$, ($\Delta \ln \mathcal{M}_c,\Delta \ln d_L)$, ($\Delta \ln \mathcal{M}_c,\Delta \Omega)$ and ($\Delta \ln d_L,\Delta \Omega)$. These choices are motivated by typical requirements for population or cosmology studies. 
For instance, to measure the Hubble parameter using Dark Sirens methods~\cite{Schutz:1986gp, DelPozzo:2011vcw, LIGOScientific:2021aug, Finke:2021aom, Demasi:2024yiv}, one would be mostly interested in the luminosity distance and the sky localization. 
While one might be interested in the detector-frame chirp mass and the luminosity distance used in the Spectral Siren method~\cite{Farr:2019twy, Gray:2023wgj, Mastrogiovanni:2021wsd, Mastrogiovanni:2023emh}

While the \texttt{2L\_45} configuration is the best one when we consider a single parameter at a time, this does not hold when we require that pairs of parameters are measured with some threshold accuracy. In this case, the differences between \texttt{2L\_0} and \texttt{2L\_45} are not significant, with the notable case of the chirp mass and the sky precision where the \texttt{2L\_0} network performs better than the \texttt{2L\_45} network. These findings suggest the need for a rigorous, more comprehensive analysis,  starting from the FIMs, to determine which network performs better given the different science cases, see e.g.~\cite{Mastrogiovanni:2023zbw, Borghi:2023opd,Gray:2023wgj, Muttoni:2023prw}.

\begin{figure}[h!]
    \centering
    \includegraphics[width=0.45\linewidth]{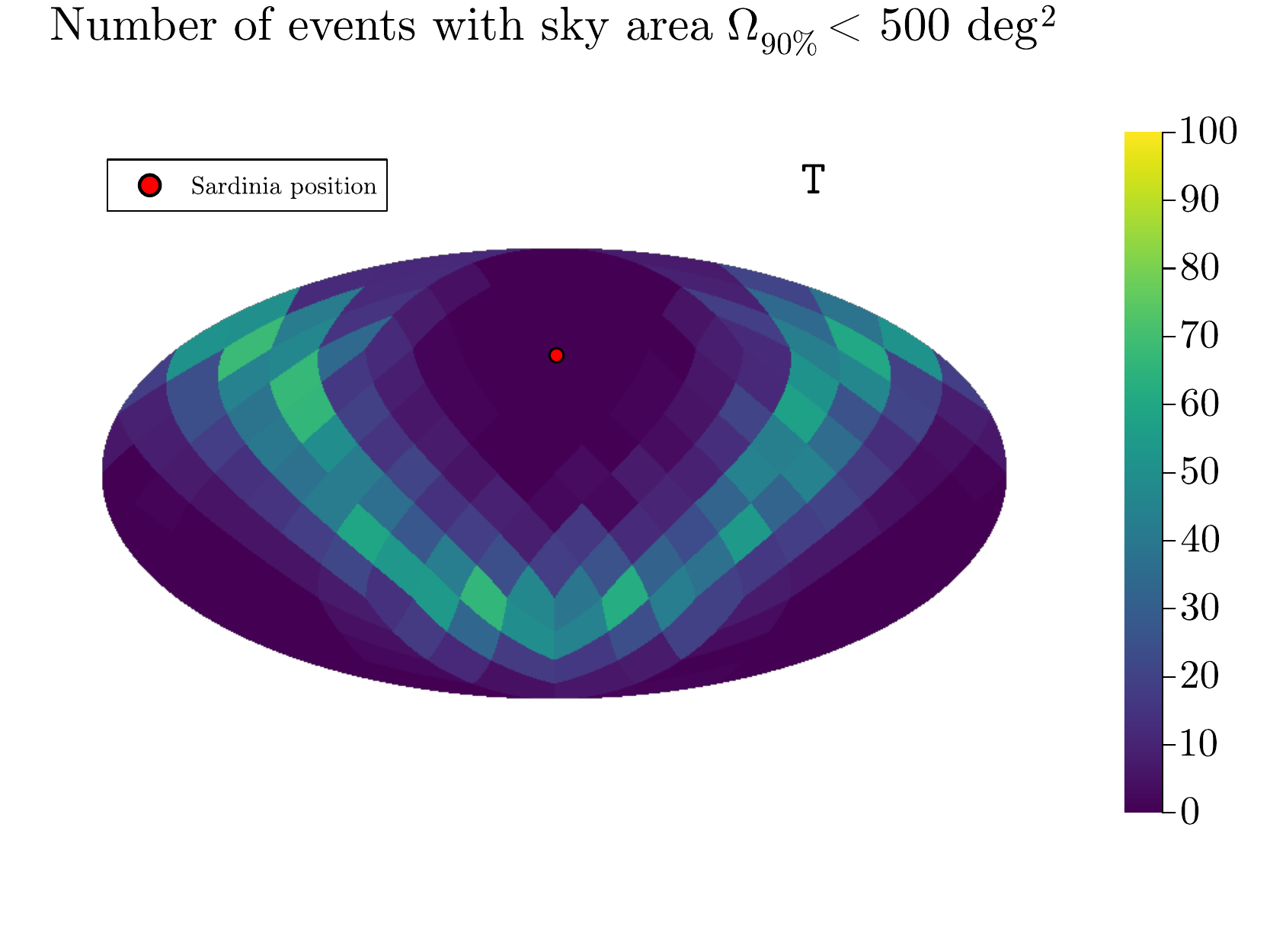}
    \includegraphics[width=0.45\linewidth]{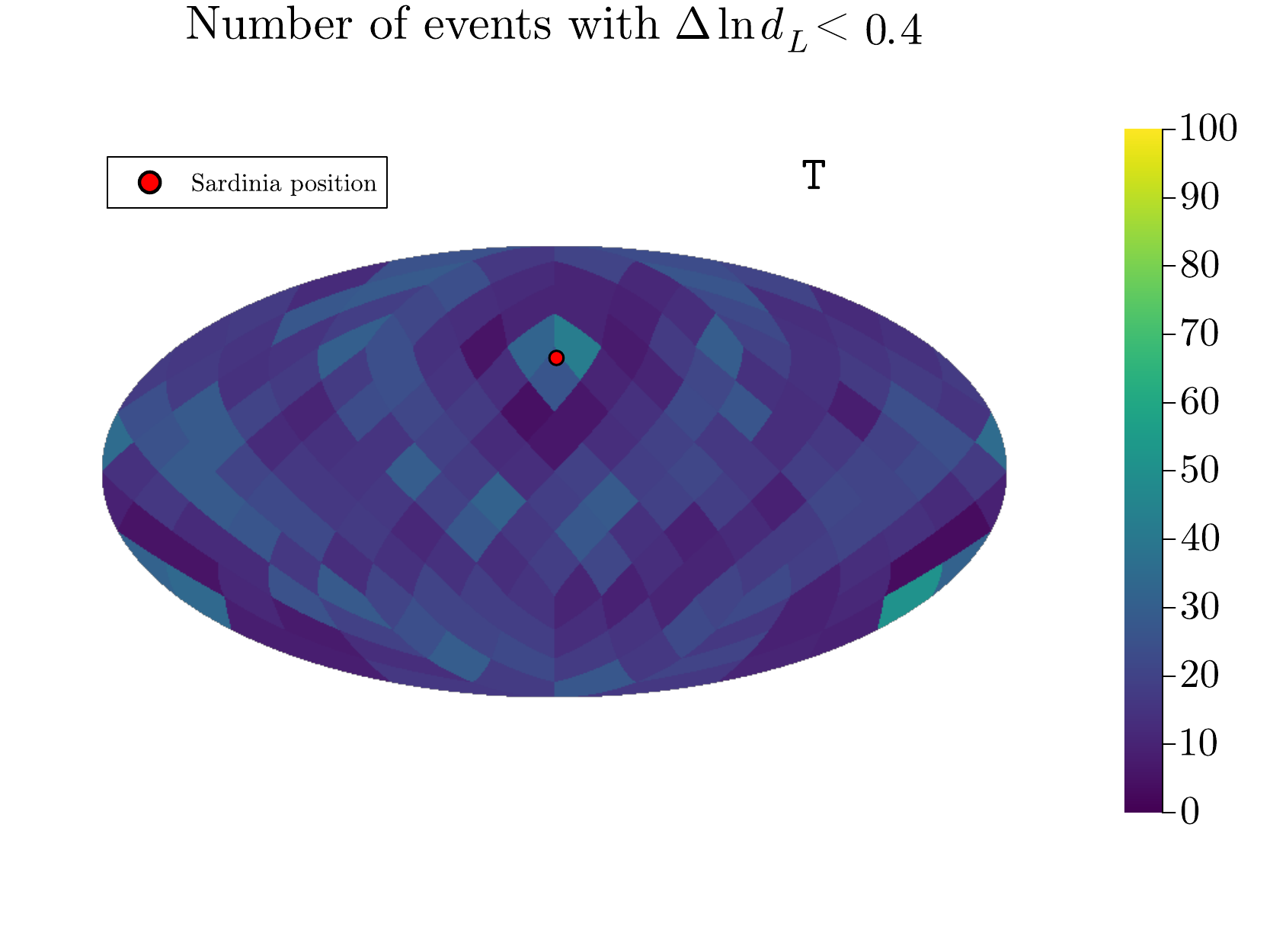}
    \includegraphics[width=0.45\linewidth]{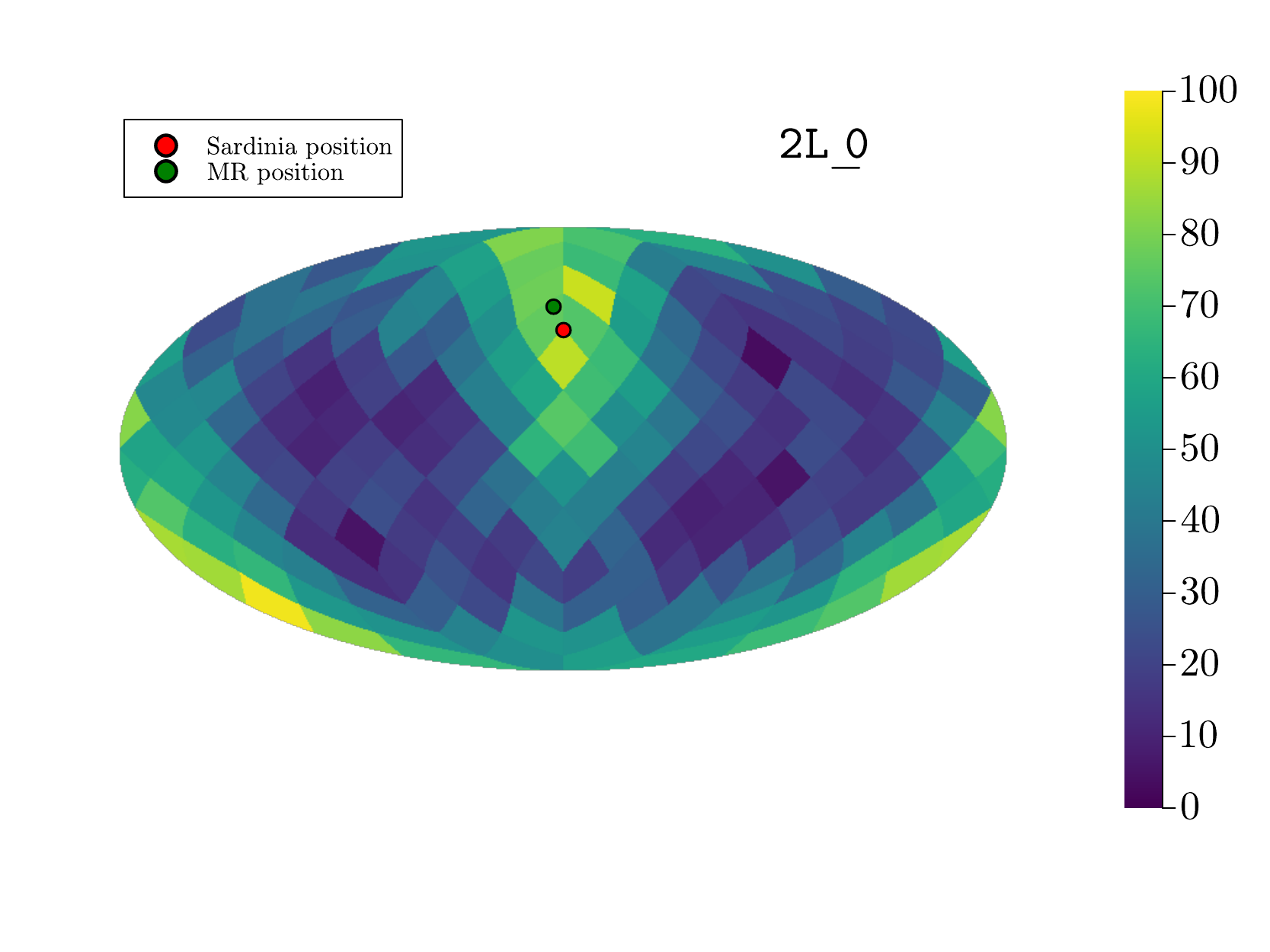}
    \includegraphics[width=0.45\linewidth]{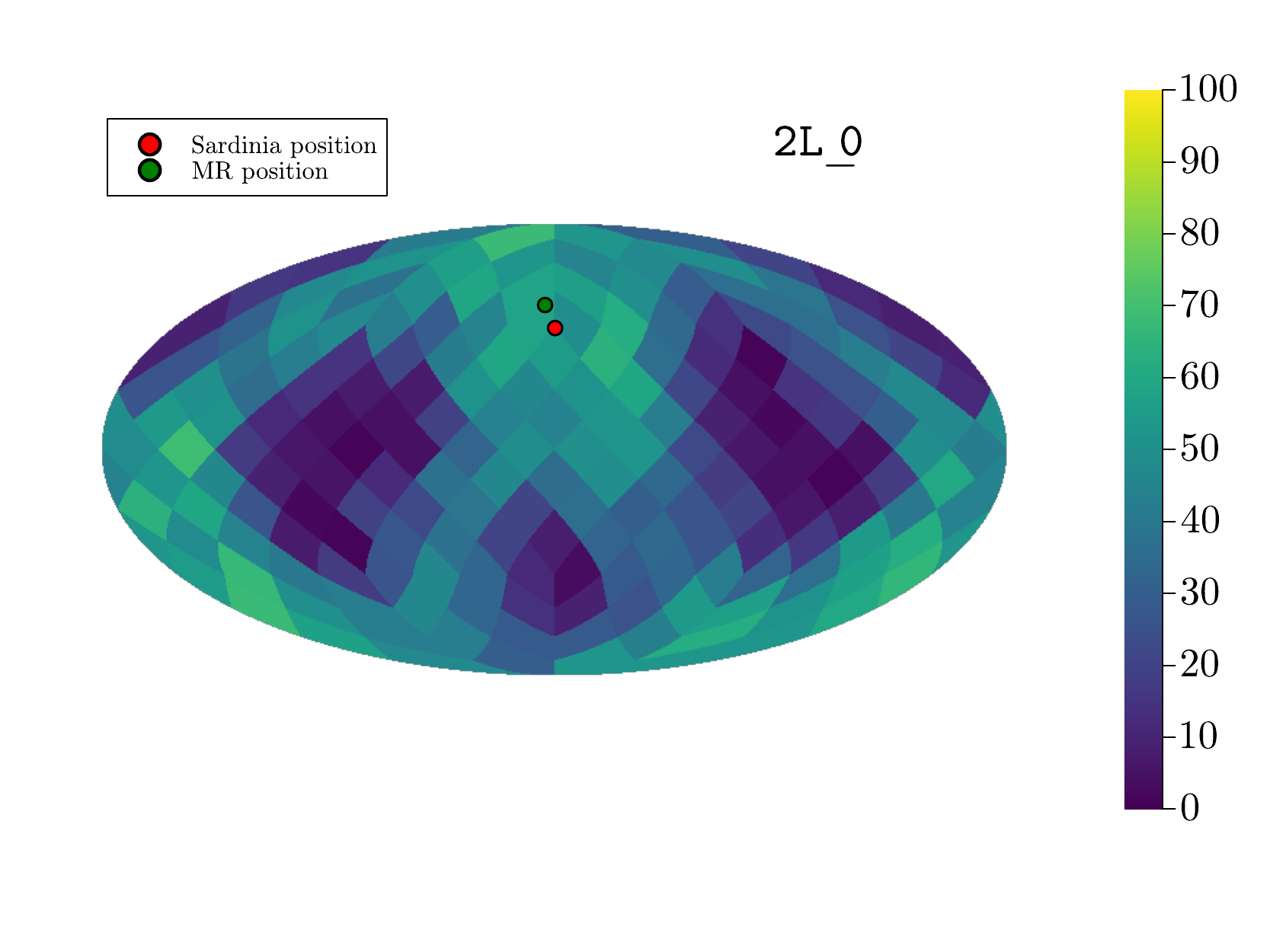}
    \includegraphics[width=0.45\linewidth]{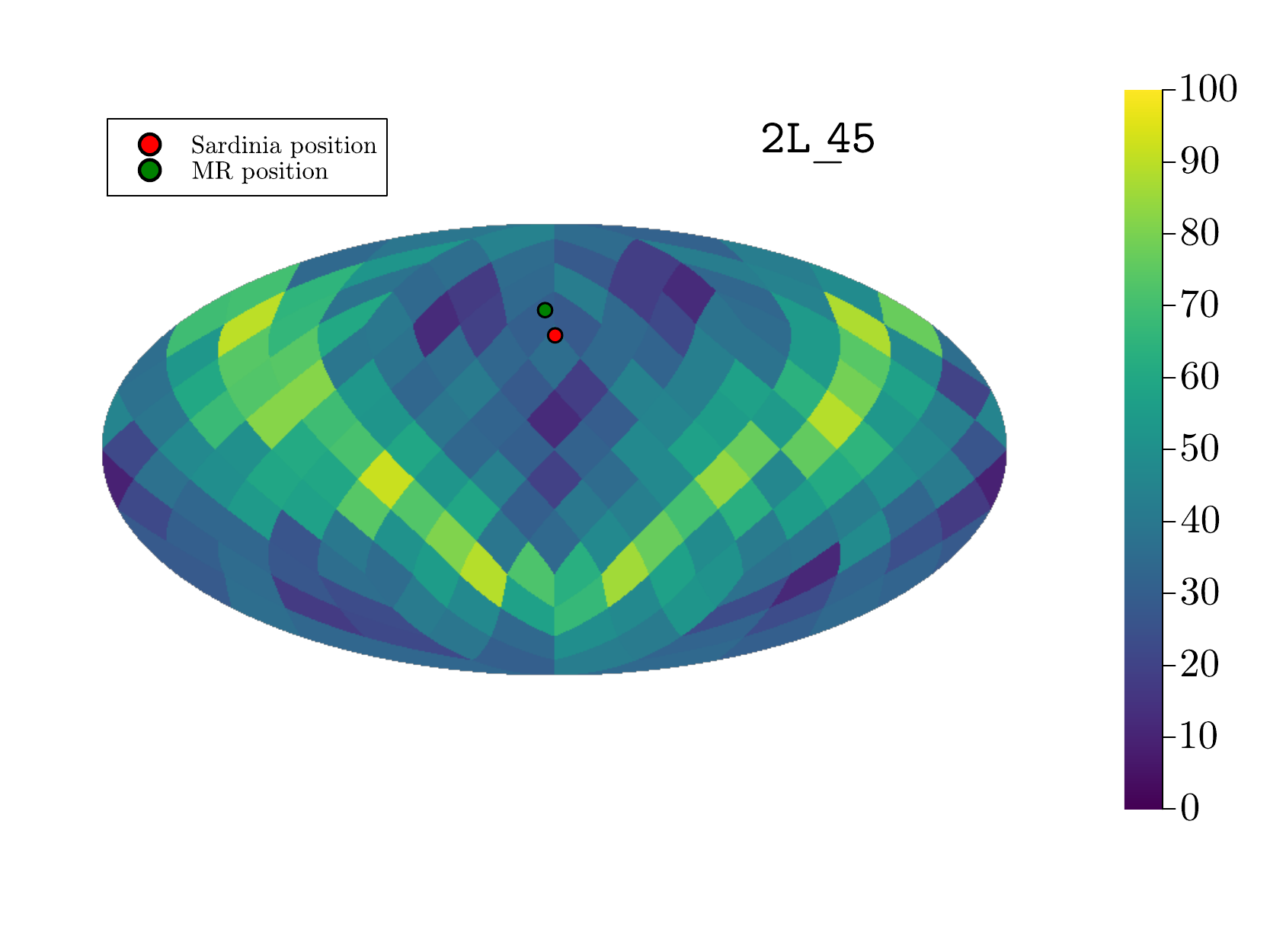}
    \includegraphics[width=0.45\linewidth]{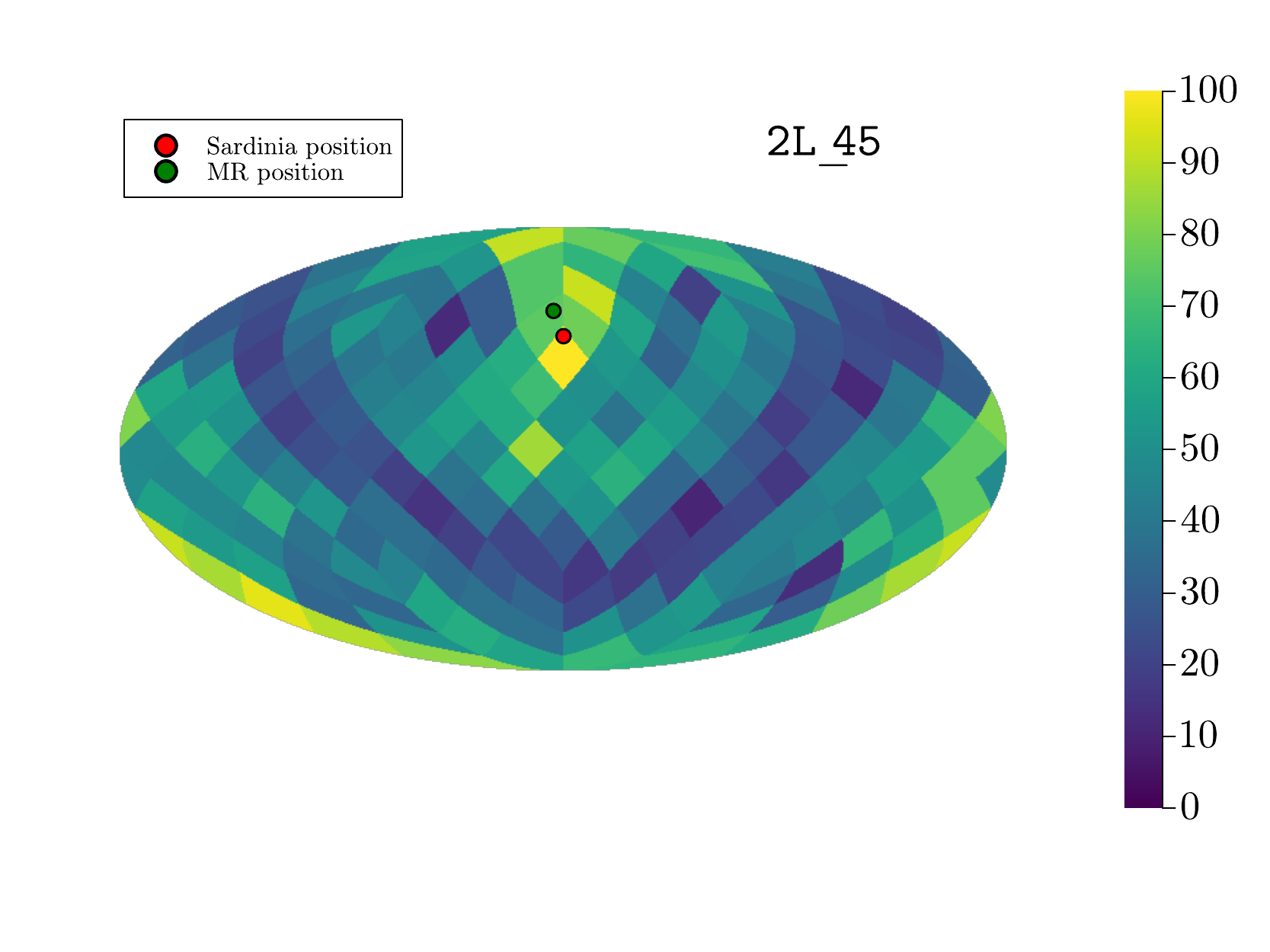}
    \caption{In Figure, we plot the sky maps as seen by the detectors, and the color coding represents the number of events with a precision better than the threshold indicated in the title. The first column shows the number of events with a precision better than 500 $\deg^2$ in sky area, while the second column shows the number of events with a relative luminosity distance precision better than 40\%. The first row shows the triangular network, the second row shows the \texttt{2L\_0} network, and the third row shows the \texttt{2L\_45} network. The red and green points represent the two detectors of each network. 
    \textit{Top:} triangular. There is a clear band of larger sensitivity for the sky area, which is measured best in the plane of the detector, while the opposite happens for the luminosity distance, leading to a few events well measured in both parameters.
    \textit{Middle:} \texttt{2L\_0}. The patterns of high sensitivity are similar in the two parameters, leading to a large number of events where both sources are well measured.
    \textit{Bottom:} \texttt{2L\_45}. The patterns of high sensitivity are, as in the case of the triangular, orthogonal to each other. This way, despite having the largest sensitivity to each parameter, the combined sensitivity is significantly reduced.}
    \label{fig:BBH_sky_maps}
\end{figure}

The sky maps shown in~\cref{fig:BBH_sky_maps} can be useful to interpret the behaviours of the different networks. Each row corresponds to a different network configuration, i.e., triangular, 2L parallel, 2L at 45° (all cryogenic). For these plots, we work in the Earth reference frame, i.e., 
the unit vectors orthogonal to the plane of the detectors are static, indicated by the red and green points, while the sky rotates around them. 
The first column shows the number of events with a precision better than 500 $\deg^2$ in sky area, while the second column shows the number of events with a relative luminosity distance precision better than 40\%. We can see how the different networks have different bands of high sensitivity in the sky (represented by the lighter colors), which clearly explain the results of~\cref{fig:BBH_golden_events}. The triangular network has a band of high sensitivity to the sky area parallel to the plane of the detector. On the other hand, the luminosity distance has an opposite pattern of high sensitivity, and this leads to few events with a good parameter estimation of both quantities. The \texttt{2L\_0} network has a similar pattern of high sensitivity for the two parameters, leading to a large number of events where both parameters are well measured. The \texttt{2L\_45} network has patterns of high sensitivity that are orthogonal to each other. This means that despite having the largest sensitivity in each parameter, the combined sensitivity is significantly reduced. This is a clear example of how the network design can affect the parameter estimation and how it is important to consider the whole science cases when designing a network of detectors.
One feature which emerges from the sky maps is the fact that the maps of the sky areas are very different in the two cases \texttt{2L\_0} and \texttt{2L\_45}. There are multiple concurring reasons for this behaviour, here we lay the principal ones. The SNR is higher in the direction orthogonal to the detectors, while the derivative of the pattern functions, i.e., the quantities entering in the FIM for $(\theta,\phi)$, are larger in the plane of the detector. Moreover, in the plane of the detector, there are reduced correlations between $\theta$ and $\phi$, which help constrain the sky area. Finally, the \texttt{2L\_0} has four zeros in the pattern functions in the plane of the detector, while this is not the case for \texttt{2L\_45}, and this limits the sensitivity of \texttt{2L\_0} in the plane of the detector. Overall, the \texttt{2L\_0} network breaks more degeneracies in the case of events orthogonal to the detectors, in particular the degeneracies with other angular parameters. We include the sky maps, analogous to \cref{fig:BBH_sky_maps}, for the chirp mass and the SNR in \cref{app:checks_fig6}, further strengthening the case discussed here.
In a more realistic setting, one should also consider the errors associated with anisotropic corrections, such as peculiar velocities, lensing and other GR corrections~\cite{Begnoni:2024tvj, Hirata:2010len, Bertacca:2017dl}. These effects would be particularly important for golden events once we include CE in the network~\cite{Pieroni:2022bbh}.

\clearpage

\subsection{Binary Neutron Stars}

\begin{figure}[t!]
    \centering
    \includegraphics[width=\linewidth]{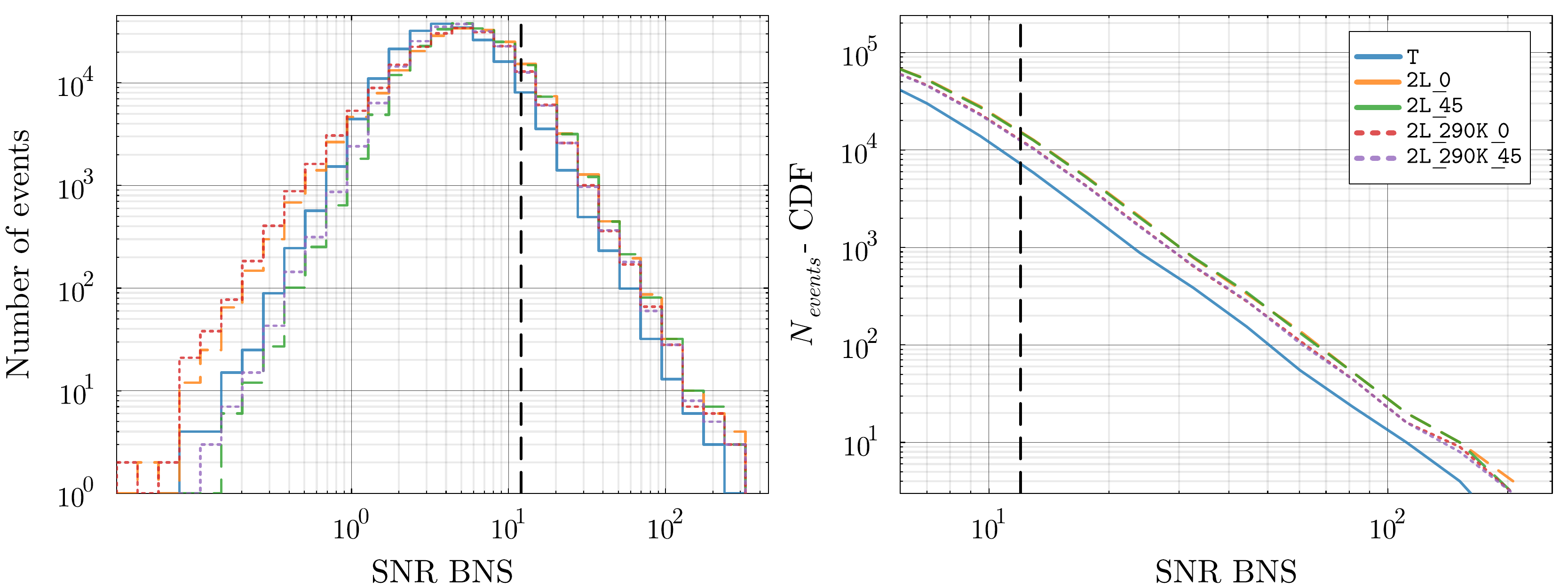}
    
    \caption{\textit{Left:} SNR histogram for the BNS events. \textit{Right:} CDF of the SNR for the same events. The vertical line represents the SNR threshold of 12. The \texttt{T} detector has the lowest SNR for detected events, while the \texttt{2L\_45} and \texttt{2L\_0} networks have the best performance.}
    \label{fig:BNS_SNR}
\end{figure}

We now analyze the BNS events using the waveform \texttt{IMRPhenomD\_NRTidalv2}~\cite{Dietrich:2019NRTidal}, and we consider 1 year of observation, for which we expect $\approx 2.2\times 10^5$ BNS events, consistent with e.g.,~\cite{Iacovelli:2022For}. We can see in~\cref{tab:BNS_SNR} and in~\cref{fig:BNS_SNR} the SNR histogram and the CDF of the SNR for the BNS events. The \texttt{T} network has a significantly worse performance, with a factor of $\approx 2$ for the number of events above a given SNR threshold. The \texttt{2L\_0} and \texttt{2L\_45} networks have a similar behaviour, with a small advantage for the \texttt{2L\_0} network. While the \texttt{2L\_290K\_0} and \texttt{2L\_290K\_45} networks are comparable, $\approx 20\%$ worse than their cryogenic counterpart. For any network configuration, at most 13.8\% of sources are detected with an SNR above 12.
\begin{table}[h!]
    \centering
    \begin{tabular}{|c|c|c|c|c|c|}
        \hline
        Network & SNR $>$ 8 & SNR $>$ 12 & SNR $>$ 20 & SNR $>$ 50 & SNR $>$ 100 \\
        \hline
        \texttt{T} & 18.2 \% & 6.9 \% & 1.6 \% & 0.1 \% & 0.1 \textperthousand \\
        \hline
        \texttt{2L\_0} & 30.8 \% & 13.8 \% & 3.6 \% & 0.2 \% & 0.3 \textperthousand \\
        \hline
        \texttt{2L\_45} & 30.6 \% & 13.5 \% & 3.5 \% & 0.2 \% & 0.3 \textperthousand \\
        \hline
        \texttt{2L\_290K\_0} & 26.9 \% & 11.5 \% & 2.9 \% & 0.2 \% & 0.2 \textperthousand \\
        \hline
        \texttt{2L\_290K\_45} & 26.7 \% & 11.3 \% & 2.8 \% & 0.2 \% & 0.2 \textperthousand \\
        \hline
        \end{tabular}
    \caption{Percentage of the SNR of BNS event above the threshold indicated in each column.}
    \label{tab:BNS_SNR}
\end{table}

\subsubsection{Fisher Information matrix analysis}
\begin{figure}[h!]
    \centering
    \includegraphics[width=\linewidth]{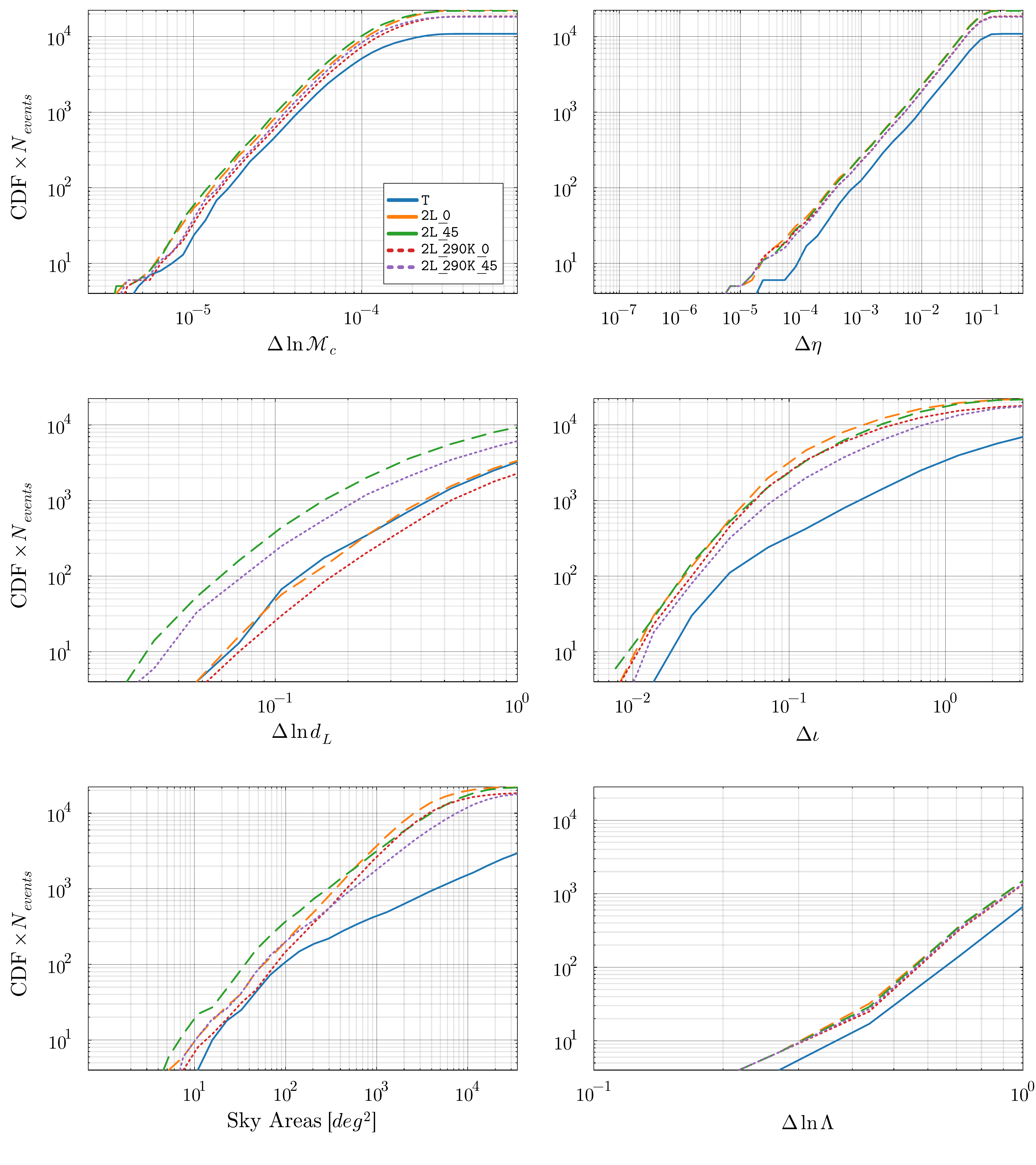}
    \caption{The parameter precisions on BNS, obtained with \texttt{IMRPhenomD\_NRTidalv2}. The triangular configuration is represented in blue, the \texttt{2L\_0} in orange dashed, the \texttt{2L\_45} in green dashed, the \texttt{2L\_290K\_0} in red dotted, and the \texttt{2L\_290K\_45} in purple dotted. \textit{Top-left}: the CDF of the parameter precisions for the relative chirp mass $\mathcal{M}_c$.
    \textit{Top-right}: the CDF of the symmetric mass ratio $\eta$. \textit{Middle-left}: the CDF of the luminosity distance $d_L$. \textit{Middle-right}: the CDF of the inclination $\iota$.
    \textit{Bottom-left}: the CDF of the 90\% sky area. \textit{Bottom-right}: the CDF of the combination of the tidal deformabilities $\tilde{\Lambda}$. As with BBH, the triangular configuration is worse than all the other configurations of a factor $\approx 1.5-2$. The \texttt{2L\_45} configuration is the best one by a slight margin. Moreover, it should also be noted that the cheaper \texttt{2L\_290K\_45} performs better than the \texttt{2L\_0} for a wide range of precisions in the luminosity distance and inclination angle.}
    \label{fig:BNS_parameters}
\end{figure}

We now analyze the parameter accuracy estimation for the BNS events. We can see in~\cref{tab:BNS_OmegaSky} and in~\cref{fig:BNS_parameters} the CDF of the parameter precisions for the BNS events. The key takeaways are similar to the BBH events, with the triangular configuration being worse than all the other configurations by a factor $\approx 1.5-2$. The \texttt{2L\_45} configuration is overall the best one for all parameters, with a significant margin only in the luminosity distance, inclination angle, and sky area. However, this is not true for all the parameter ranges of the CDF. We also plot the relative errors on $\tilde{\Lambda}$, while the constraints on $\delta\Lambda$ are uninformative.
\clearpage
\begin{table}[h!]
    \centering
    \begin{tabular}{|c|c|c|c|}
        \hline
        Network &   $\Omega$ $<$ 100 deg$^2$ & $\Omega$ $<$ 1000 deg$^2$ & $\Omega$ $<$ Whole sky  \\
        \hline
        \texttt{T} & 0.05 \% & 0.24 \% & 1.56 \% \\
        \texttt{2L\_0} & 0.08 \% & 1.48 \% & 13.26 \% \\
        \texttt{2L\_45} & 0.16 \% & 1.3 \% & 12.85 \% \\
        \texttt{2L\_290K\_0} & 0.06 \% & 1.04 \% & 10.87 \% \\
        \texttt{2L\_290K\_45} & 0.09 \% & 0.75 \% & 10.01 \% \\
        \hline
        \end{tabular}

    \caption{Percentage of the BNS events with a sky area below the threshold indicated in each column. 100\% represents the whole sources in the catalog (i.e., detected and not detected)}
    \label{tab:BNS_OmegaSky}
\end{table}
    

\subsection{Neutron Stars - Black Holes}
\begin{figure}
    \centering
    \includegraphics[width=\linewidth]{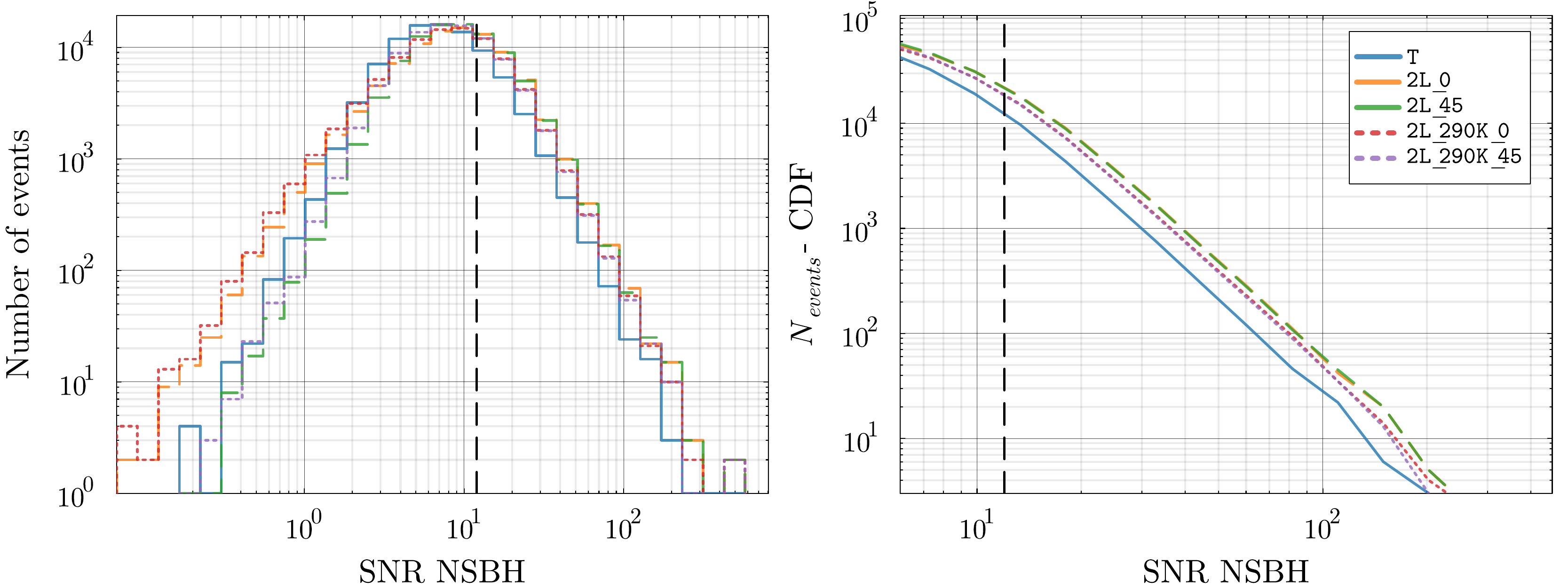} 
    \caption{\textit{Left}: SNR histogram for the NSBH events. \textit{Right}: CDF of the SNR for the same events. The vertical line represents the SNR threshold of 12. As before, the \texttt{T} detector has a significantly worse performance, while the \texttt{2L\_45} and \texttt{2L\_0} networks obtain larger overall SNRs.}
    \label{fig:NSBH_SNR}
\end{figure}

\begin{figure}
    \centering
    \includegraphics[width=\linewidth]{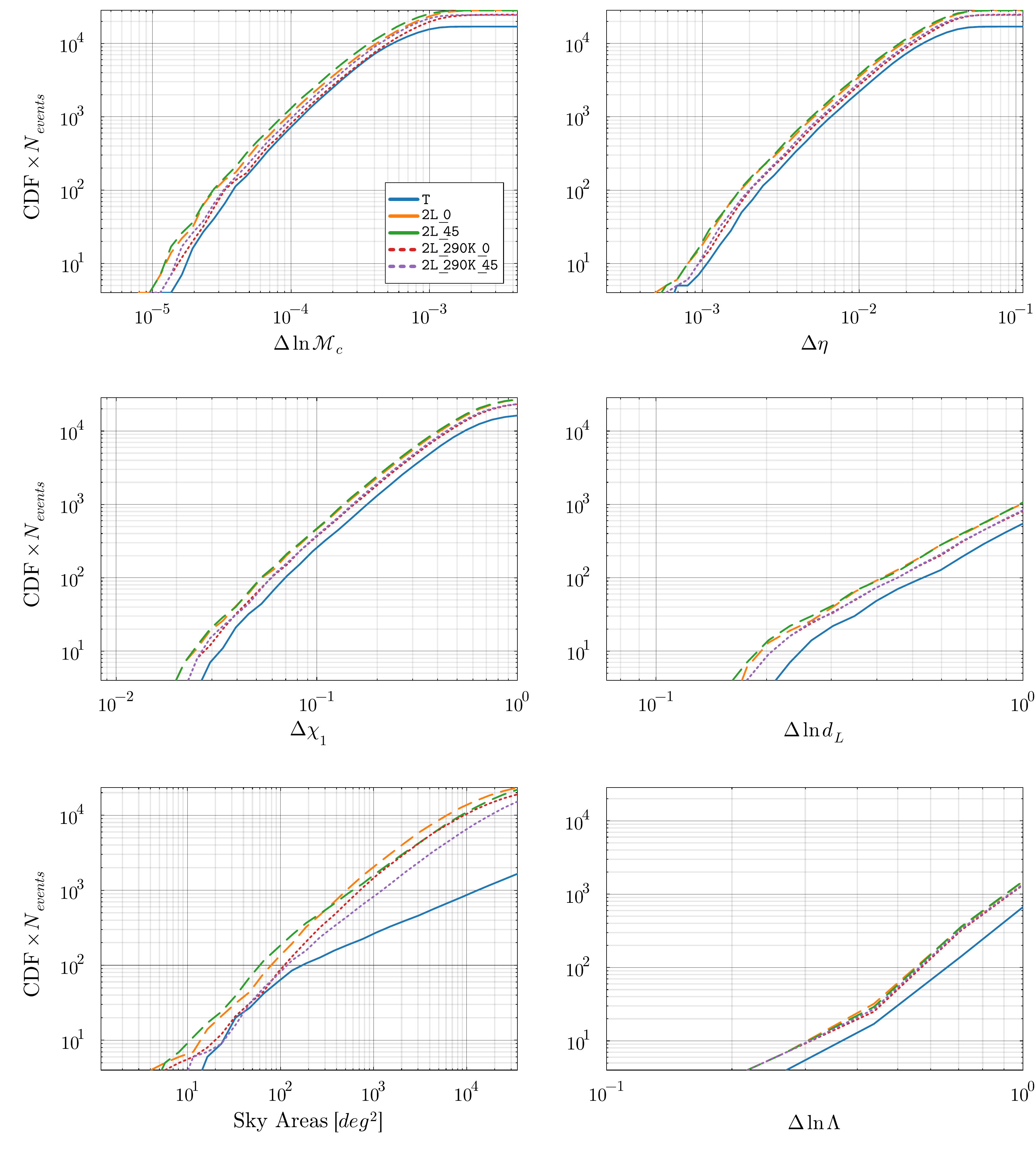}
    
    \caption{The parameter precisions of NSBH, obtained with \texttt{IMRPhenomNSBH}. The triangular configuration is represented in blue, the \texttt{2L\_0} in orange dashed, the \texttt{2L\_45} in green dashed, the \texttt{2L\_290K\_0} in red dotted, and the \texttt{2L\_290K\_45} in purple dotted. \textit{Top-left}: the CDF of the parameter precisions for the relative chirp mass $\mathcal{M}_c$.
    \textit{Top-right}: the CDF of the symmetric mass ratio $\eta$. \textit{Middle-left}: the CDF of the first spin $\chi_1$. \textit{Middle-right}: the CDF of the relative luminosity distance $d_L$.
    \textit{Bottom-left}: the CDF of the 90\% sky area $\Omega_{90\%}$. \textit{Bottom-right}: the CDF of the tidal deformability $\tilde\Lambda$. 
    We recover the same hierarchy as for BBH and BNS.}
    \label{fig:NSBH_parameters_12}
\end{figure}

NSBHs are analyzed using the \texttt{IMRPhenomNSBH} waveform, and we consider one year of observations, for which we expect $ N\approx8.9\times 10^4$. As we can see from~\cref{tab:NSBH_SNR} and~\cref{fig:NSBH_SNR}, the \texttt{2L} networks detect approximately 30\% of the sources, while the triangular configuration stops at 20\%. We can also look at the sky area as presented in~\cref{tab:NSBH_OmegaSky} and~\cref{fig:NSBH_parameters_12}. We can see that the triangular design fails to make good constraints on the sky positions for most of the sources. While as for the case of BBH, the highest share of precise events, i.e., under 100 $\deg^2$, is largest for \texttt{2L\_45} while the largest share of events under 1000 $\deg^2$ is \texttt{2L\_0}. Highlighting also for NSBH a behaviour seen in both BNS and NSBH.
\begin{table}[h!]
    \centering
    \begin{tabular}{|c|c|c|c|c|c|}
        \hline
        Network & SNR $>$ 8 & SNR $>$ 12 & SNR $>$ 20 & SNR $>$ 50 & SNR $>$ 100 \\
        \hline
        \texttt{T} & 39.6 \% & 19.1 \% & 5.4 \% & 0.4 \% & 0.4 \textperthousand \\
        \hline
        \texttt{2L\_0} & 54.5 \% & 32.1 \% & 11.1 \% & 0.8 \% & 1.1 \textperthousand \\
        \hline
        \texttt{2L\_45} & 55.9 \% & 31.7 \% & 10.9 \% & 0.8 \% & 1.0 \textperthousand \\
        \hline
        \texttt{2L\_290K\_0} & 49.9 \% & 27.9 \% & 9.1 \% & 0.7 \% & 0.9 \textperthousand \\
        \hline
        \texttt{2L\_290K\_45} & 50.8 \% & 27.5 \% & 8.9 \% & 0.6 \% & 0.8 \textperthousand \\
        \hline
        \end{tabular}

    \caption{Percentage of the SNR of NSBH event above the threshold indicated in each column.}
    \label{tab:NSBH_SNR}
\end{table}

\begin{table}[h!]
    \centering
    \begin{tabular}{|c|c|c|c|}
        \hline
        Network &   $\Omega$ $<$ 100 deg$^2$ & $\Omega$ $<$ 1000 deg$^2$ & $\Omega$ $<$ Whole sky  \\
        \hline
        \texttt{T} & 0.06 \% & 0.28 \% & 1.87 \% \\
        \texttt{2L\_0} & 0.12 \% & 1.98 \% & 26.31 \% \\
        \texttt{2L\_45} & 0.18 \% & 1.52 \% & 24.73 \% \\
        \texttt{2L\_290K\_0} & 0.08 \% & 1.37 \% & 21.39 \% \\
        \texttt{2L\_290K\_45} & 0.07 \% & 0.8 \% & 17.17 \% \\
        \hline
        \end{tabular}
    \caption{Percentage of the NSBH events with a sky area below the threshold indicated in each column. 100\% represents the whole sources in the catalog.}
    \label{tab:NSBH_OmegaSky}
\end{table}
An important remark: the \texttt{IMRPhenomNSBH} waveform is not calibrated for $\chi_2\ne 0$. However, we expect this would lead to minor differences since the important parameter in the waveform is the spin weighted on the mass of the two objects, where the NS is significantly lighter than the BH.
The results we find closely resemble the ones from BNS, because no higher harmonics are present in the waveforms of both sources.

    

    

\section{Hamiltonian Monte Carlo}\label{sec:MCMC}
\begin{figure}
    \centering
    \includegraphics[width=\linewidth]{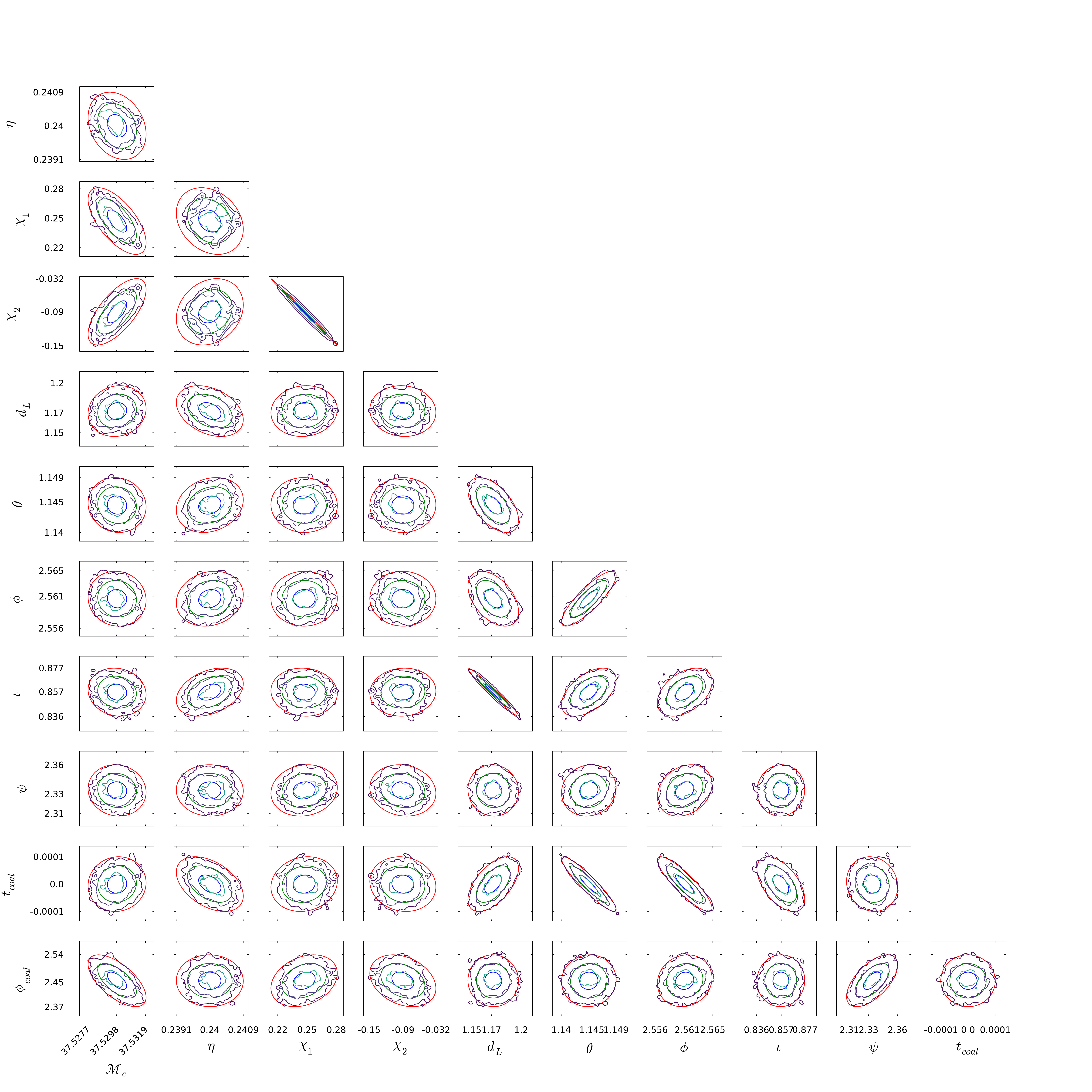}
    \caption{The 2d contours for all the couples of parameters for a BBH event. The smooth ellipses show the 1, 2 and 3 $\sigma$ intervals predicted by the FIM, respectively in blue, green and red. The irregular contours, indicated with shades from cyan to purple, represent instead the 1, 2 and 3 $\sigma$ intervals sampled by the HMC. The FIM and HMC estimates agree remarkably well for most parameters, with the sampler accurately capturing even strong degeneracies. Discrepancies arise in the spin parameters, where we expect significant non-Gaussian features.}
    \label{fig:HMC}
\end{figure}

We now present in~\cref{fig:HMC} a comparison of the FIM result with a full Bayesian analysis. We perform this test under many simplifying assumptions, and using a new sampler we are currently developing. For our study, we consider a realization without scatter, i.e, the posterior peak is at the injected event position. This translates to injecting $n(t) = 0$ in~\cref{eq:signal}. Moreover, we choose an event with large SNR (SNR = 800), to make the posterior more Gaussian (see e.g.~\cite{Vallisneri:2007ev}) and we use a network of 3 detectors, the network \texttt{2L\_45} plus a 40 km Cosmic Explorer~\cite{Reitze:2019CE, Evans:2021CE} in the US.
Under such assumptions, we apply an Hamiltonian Monte Carlo (HMC)~\cite{Neal:2011HMC} directly to the event posterior. An application of HMC to CBC has been recenlty proposed in~\cite{Perret:2025wox}. HMC is very well suited for sampling $\mathcal{O}(10)$ parameter functions, when gradients evaluations are computationally cheap, even in the presence of strong degeneracies.  This makes it particularly well suited to gravitational-wave event posteriors, where automatic differentiation can be used to compute gradients efficiently. One of the main caveats, apart from many technical difficulties, is that the HMC is not suited for exploring very multimodal distributions~\cite{Neal:2011HMC}. This can be fixed in our case; Since we are simulating future events, we already know the exact location of the peak in the zero-noise case. This assumption can be relaxed even in the case of a proper noise injection, because the peak will not move by a large amount. A key advantage of the HMC is that it can be guided by the curvature at the posterior peak, i.e., the second derivative of the log-posterior~\cite{Neal:2011HMC}, which is well approximated by the FIM~\cref{eq:FIM} Notice that the FIM also implements the ensemble average, of the second derivative of the log-posterior, over the noise realizations. When the true posterior is nearly Gaussian, this curvature information makes sampling significantly more efficient. 

Here we examine a very high SNR case, while a more complete and realistic analysis will be performed in a future work. The main advantage is the extremely cheap computational cost of this sampling. The chain we present here has 15000 steps, uses a 3 L-shaped detectors with \texttt{IMRPhenomXHM} and took only one CPU hour to run. From the corner plot in~\cref{fig:HMC}, we can see a good overlap between the FIM results (smooth curves) and the samples (irregular curves). The largest differences emerge in the intrinsic parameters, in particular the spins, where large non-gaussianities are expected. Most of the 2d contours are in very good agreement with the FIM result.
Across all intrinsic and extrinsic parameters, the FIM predicted $68\%$ confidence regions (blue) and the Hamiltonian Monte Carlo posteriors (cyan) overlap almost exactly, with discrepancies in the one‐sigma widths of less  $10\%$ for mass and distance parameters. For example, the chirp mass uncertainty is around $\sigma(\mathcal{M}_{c})\simeq 2\times 10^{-3} M_{\odot}$ in both methods, and the symmetric mass-ratio width agrees to better than  $5 \%$. 
Among all parameters, the largest differences appear in the mass ratio and in the spin amplitudes, which were expected given the known non-Gaussianity of these posteriors~\cite{Ashton:2021anp, KAGRA:2021TC3}, but overall, the two approaches yield quantitatively similar uncertainty estimates. Possible reasons for the differences in the intrinsic parameters include the fact that our sampler is informed by the FIM, making it more difficult to sample in regions where the curvature is different from the expected one. However, a better analysis of the performance of this sampler would require a P-P plot study, which is beyond the scope of this analysis.

\section{Conclusions}\label{sec:conclusions}
In this work, we forecast the detectability and the uncertainties on the relevant parameters for BBH, BNS, and NSBH using different ET configurations. We have considered a triangular configuration, and four different 2L networks, with two different orientations between the two L's and with and without the cryogenic technology, which we named: \texttt{T}, \texttt{2L\_0}, \texttt{2L\_45}, \texttt{2L\_290K\_0}, \texttt{2L\_290K\_45}. 
Our results confirm that the best network when looking at the different parameter precisions is the \texttt{2L\_45} by a slight margin over the \texttt{2L\_0} network, with \texttt{2L\_290K\_45} and \texttt{2L\_290K\_0} following. These results are valid for most of the parameters and ranges of the parameter uncertainties.

\vspace{0.3cm}
The situation changes when we examine pairs of parameters, requiring both accuracies to respect some requirements. This situation resembles the real science cases where we are usually interested in more than one parameter, e.g., to measure the Hubble constant, both the sky position and luminosity distance are required. In this more realistic situation, \texttt{2L\_45} and \texttt{2L\_0} lead to very similar performances, requiring more refined analysis.
Our findings indicate that orientation has only a limited impact on the uncertainty estimates, as we obtain comparable results regardless of the chosen orientation. Our analysis remarks that choosing the final ET configuration must be guided by specific scientific targets. Moreover, a complete analysis would also require the inclusion of the stochastic gravitational wave background, since the detector orientation significantly influences its detection, together with an optimal data-analysis strategy~\cite{Franciolini:2025leq, Caporali:2025mum, Cireddu:2023ssf}.

\vspace{0.3cm}
To perform this work analysis, we developed the open source code \texttt{GWJulia}~\github{https://github.com/andrea-begnoni/GW.jl}. We also performed a proof of concept Hamiltonian Monte Carlo analysis by leveraging the \texttt{Julia} waveform implementations in \texttt{GWJulia}. This analysis proves effective and very computationally efficient, opening the window for more realistic analysis, which could significantly reduce the computational burden of parameter estimation of CBC.

\section*{Acknowledgments}
The authors would like to thank Cecilio García-Quirós for helping with \texttt{IMRPhenomXHM} and M. Mancarella and F. Iacovelli for useful discussions. Moreover, the authors would like to thank the anonymous referee for useful comments on previous versions of the paper. A.Ri. would like to thank W. Del Pozzo for helpful discussions and feedback.
A.B. and A.Re. are supported by ICSC – Centro Nazionale di Ricerca in High Performance Computing, Big Data and Quantum Computing, funded by European Union – NextGenerationEU”. 
M.P. acknowledges the hospitality of Imperial College London, which provided office space during some parts of this project. The work of M.P. is supported by the Comunidad de Madrid under the Programa de Atracción de Talento Investigador with number 2024-T1TEC-31343. We thank INFN for the support.

\appendix
\crefalias{section}{Appendix}

\section{Waveforms}\label{app:waveform}
Here we summarize the characteristics of some waveforms present in the literature. Since the code uses automatic differentiation, the waveforms had to be rewritten in Julia and can be found in the repository.

\begin{itemize}
\item $\texttt{IMRPhenomD}$~\cite{Husa:2015PhD, Khan:2015PhD}: one of the most widespread waveforms, now outdated due to the lack of higher modes, which break some parameter degeneracies. It approximates the signals coming from BBH mergers, with non-precessing spins up to $|\chi_z| = 0.85$ and mass ratios up to $q = m_1/m_2 = 18$.

\item $\texttt{IMRPhenomXAS}$~\cite{Pratten:2020XAS}: an improved version of $\texttt{IMRPhenomD}$. It includes modifications of the phenomenological ansatz and uses a larger set of numerical relativity simulations for the calibration. Moreover, it uses a systematic approach to modeling the dependence of phenomenological parameters, and it is calibrated for mass ratios $q$ up to 1000 into the calibration dataset and spins up to $|\chi_z| = 0.9$.

\item $\texttt{IMRPhenomHM}$~\cite{London:2017HM, Kalaghatgi:2019HM}: a full waveform model, which takes into account not only the quadrupole of the signal but also the subdominant multipoles ($\ell$, $m$) = (2, 1), (3, 2), (3, 3), (4, 3) and (4,4), that can be particularly relevant to better describe the signal coming from BBH systems. 
The calibration range is the same as the $\texttt{IMRPhenomD}$. Notice that each multipole is proportional to a different spherical harmonic $_{-2}Y^{\ell m}$~\cite{Goldberg:1966SpH}, with a specific dependence on the inclination angle $\iota$. Thus, higher modes meaningfully impact parameter estimations since they help to break the degeneracy between $d_L$ and $\iota$, which is present for waveforms including only the fundamental mode. In this waveform model, a map was created to relate the fundamental mode to the subdominant harmonics. This map is based on approximate scaling relations, meaning that only the fundamental mode is calibrated to numerical data.

\item $\texttt{IMRPhenomXHM}$~\cite{Garcia-Quiros:2020XHM}: this waveform improves on $\texttt{IMRPhenomHM}$ as it extends the approach of \texttt{IMRPhenomXAS} on higher order harmonics. Moreover, it calibrates each mode used in the model, i.e., (2, 1), (3, 2), (3, 3), and (4,4). With the important inclusion of mode mixing of the mode (3, 2) with the fundamental. It is restricted to the quasi-circular (i.e., non-eccentric) and nonprecessing part of the parameter space (same as \texttt{IMRPhenomHM}). It is calibrated, as $\texttt{IMRPhenomXAS}$ for mass ratios up to $q= 1000$ and spins up to $|\chi_z|= 0.9$.

\item $\texttt{IMRPhenomXPHM}$~\cite{Pratten:2020XPHM}: an extension of $\texttt{IMRPhenomXHM}$ to the precessing case. It represents the state-of-the-art of \texttt{Phenom} waveforms for what concerns waveforms with no tidal deformability (i.e., BBH) and it is the standard waveform currently used by the LVK collaboration, e.g.,~\cite{LIGOScientific:2024NSBH}. It is not present in $\texttt{GWJulia}$.

\item $\texttt{IMRPhenomD\_NRTidalv2}$~\cite{Dietrich:2019NRTidal}: tidal extension of $\texttt{IMRPhenomD}$, allowing for the Love number $\Lambda$ to be different from 0. It includes tidal corrections not only in the phase but also in the amplitude. It also includes quadrupole and octupole tidal deformability terms, which couple to the spins of the binary up to 3.5 PN order. In the calibration, it uses different equations of state. There is also the presence of a taper to terminate the waveform after the merger. The waveform has been tested for $M_{1/2}\in [1, 3] M_{\odot}$, spins up to $|\chi_z| = 0.6$, and tidal deformability up to $\Lambda = 5000$. At the time of writing, a new version has been released $\texttt{IMRPhenomD\_NRTidalv3}$~\cite{Abac:2023Tv3}, which is not present in $\texttt{GWJulia}$.

\item $\texttt{IMRPhenomNSBH}$~\cite{Pannarale:2015NSBH, Thompson:2020PhenomNSBH}: it is based for the inspiral on $\texttt{IMRPhenomC}$~\cite{Santamaria:2010PhenomC}, which is a BBH waveform. $\texttt{IMRPhenomNSBH}$ adds the tidal effects due to the merger by considering the deformability of the NS. The merger can be divided into four types: “disruptive,” “nondisruptive,” and “mildly disruptive” with and without a torus remnant, which lead to different phase and amplitude evolutions. 
One should also note that the spin of the secondary object is fixed to zero, while the spin of the primary is free to vary up to $|\chi_z| = 0.5$. The waveform is calibrated for mass ratios up to $q = 15$.

\item $\texttt{TaylorF2\_RestrictedPN}$~\cite{Buonanno:2009F2, Pan:2007F2, Boyle:2009F2}: it is an inspiral-only waveform and it is not calibrated with NR simulations. This makes it suited to test a large variety of mergers (i.e., NS outside their mass window, or compact objects with very high spins), with the caveat that the results in this case will overestimate the errors due to the lack of information from the merger and ringdown. However, this will not be a problem for light CBC where the merger falls outside the PSD of the detector. It can be used in all the parameter space for BBH, BNS, and NSBH, since there is also a tidal extension~\cite{Wade:2014Tidal}, which enters at PN order 5 and 6. Moreover, we implemented the eccentric extension~\cite{Moore:2016ecc} up to 3PN order. There is no limitation in the parameter range, except the eccentricity which is limited to $e\approx 0.1$.

\end{itemize}

\section{Additional checks on the golden events sky maps}\label{app:checks_fig6}

\begin{figure}
    \centering
    \includegraphics[width=0.5\linewidth]{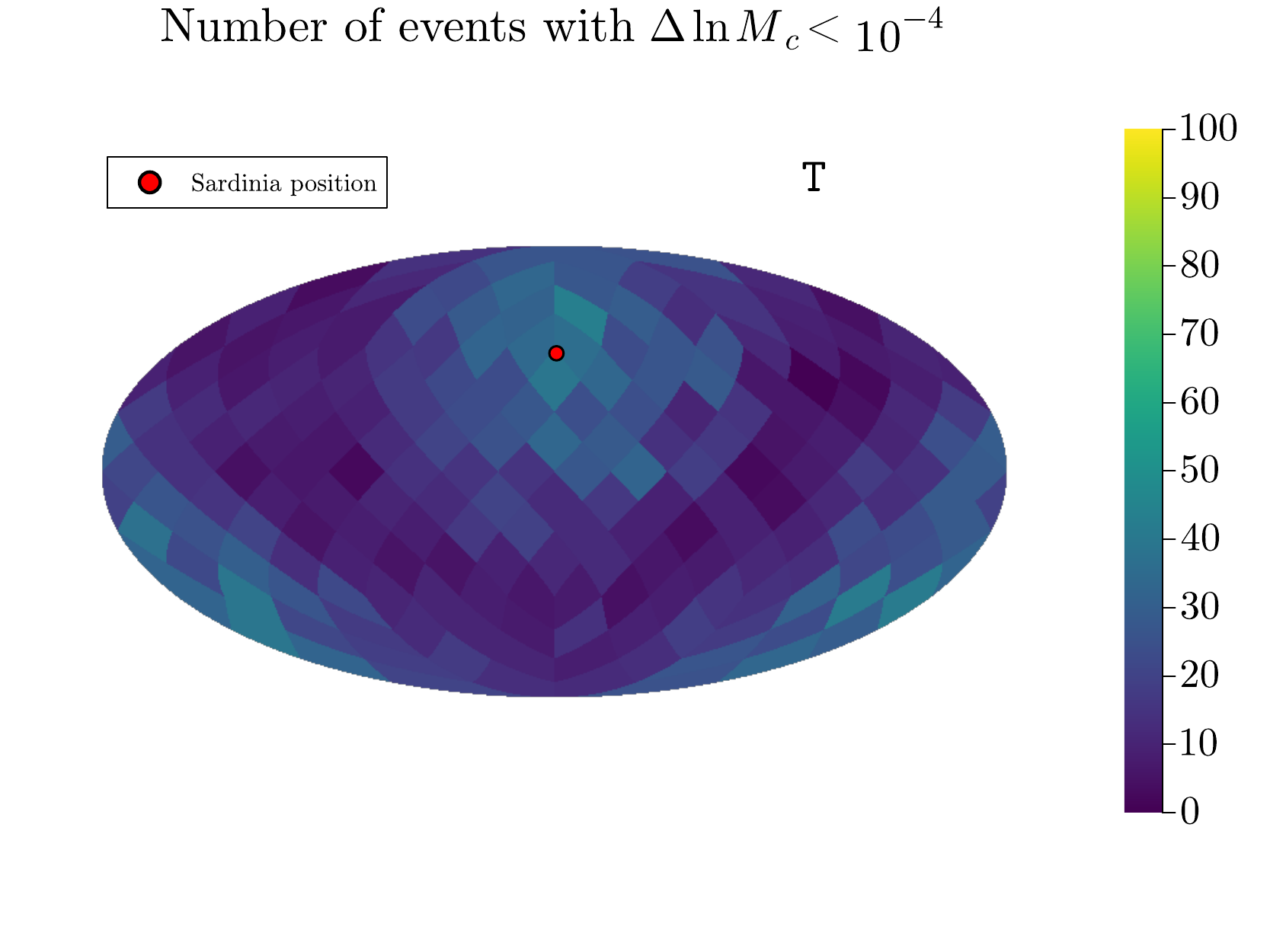}
    \includegraphics[width=0.49\linewidth]{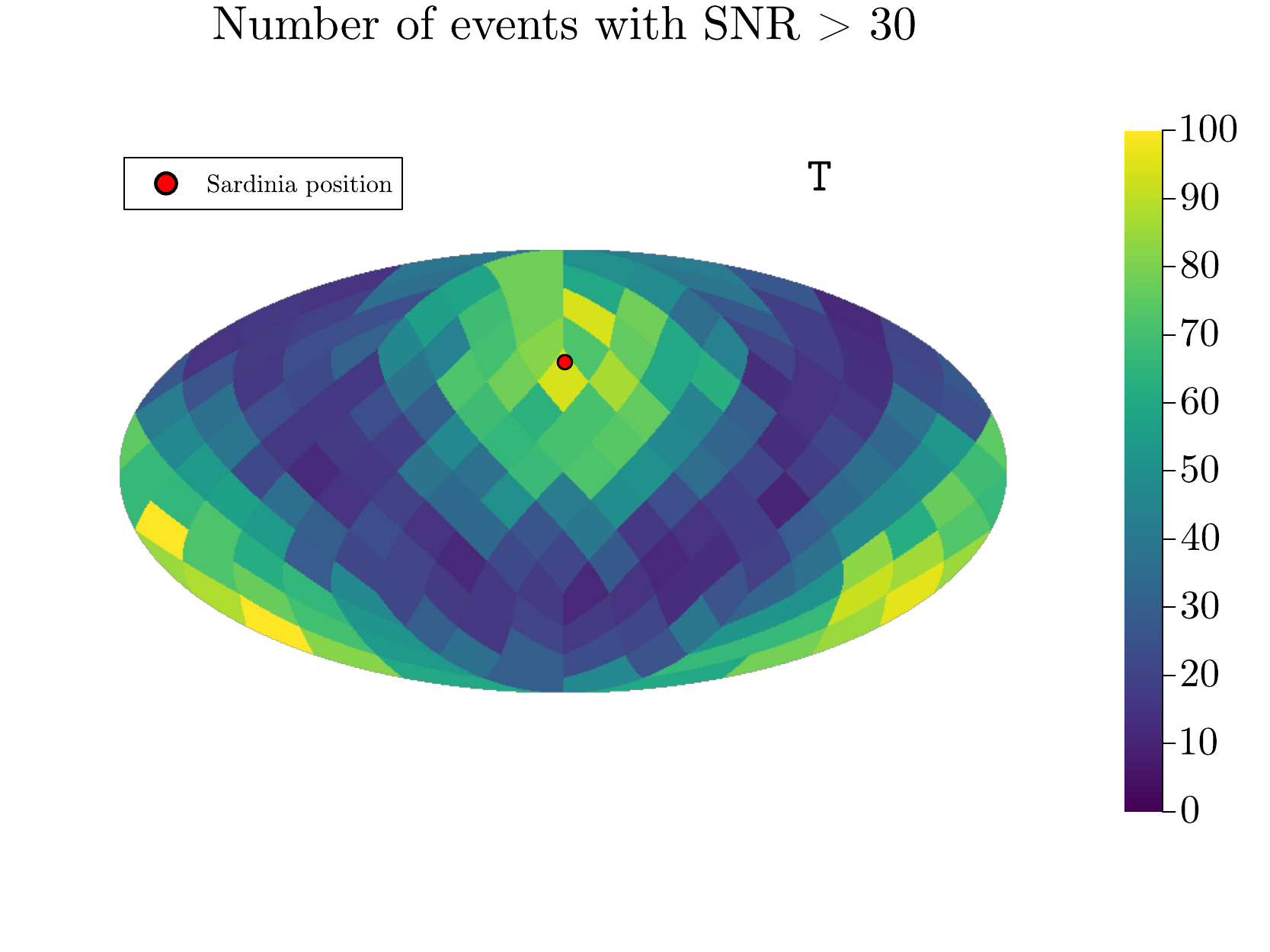} 
     \includegraphics[width=0.5\linewidth]{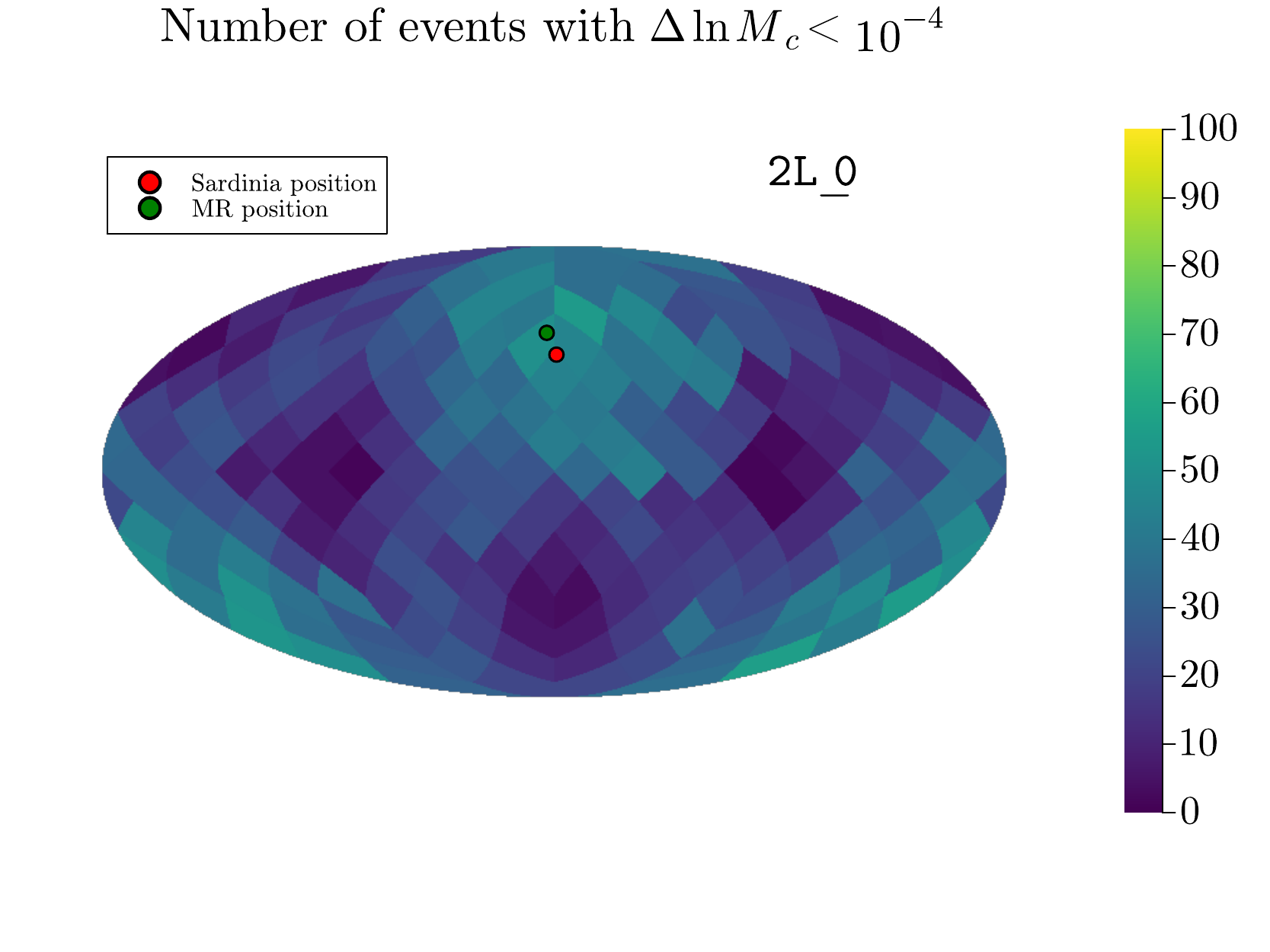}
    \includegraphics[width=0.49\linewidth]{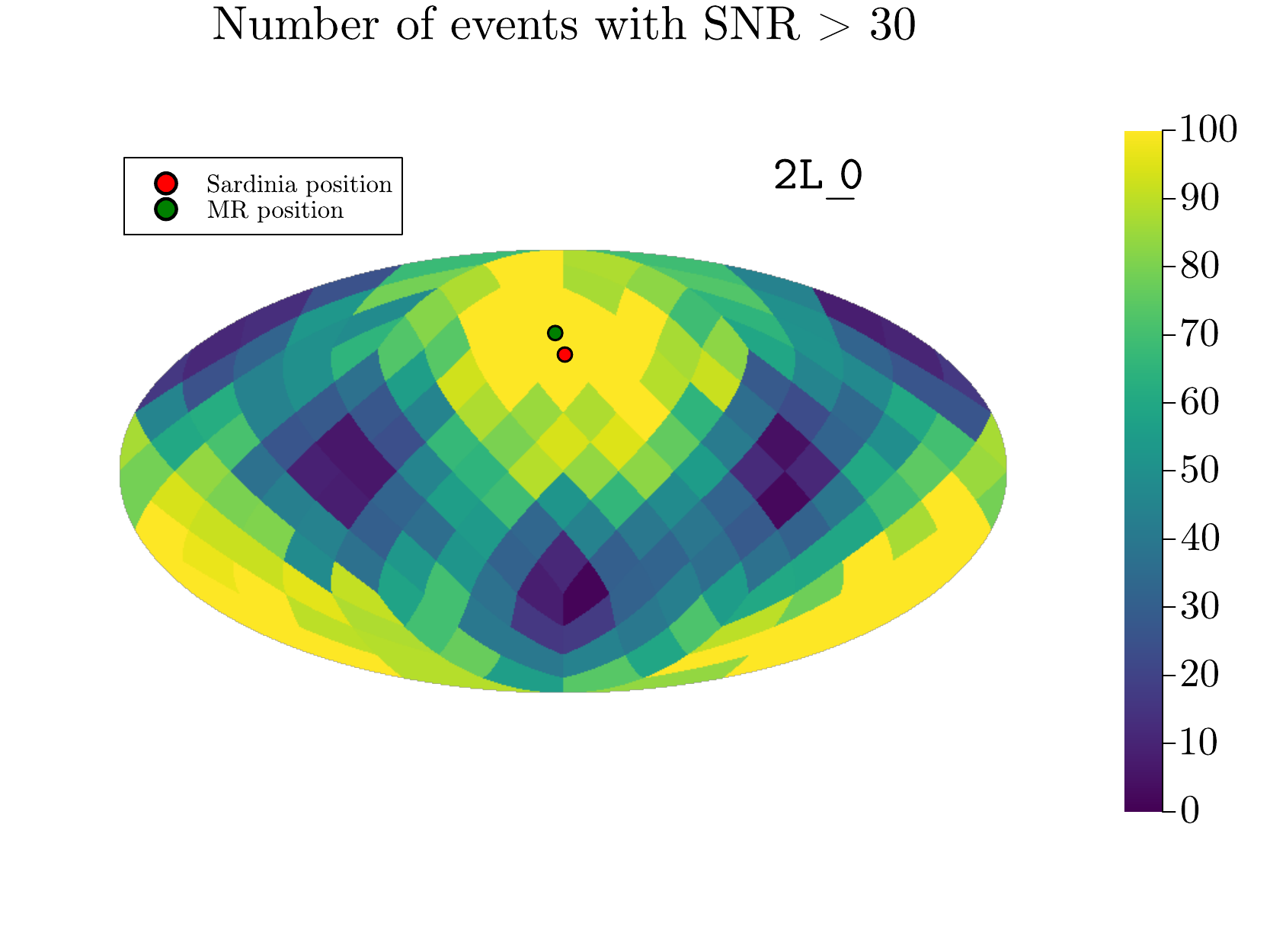} 
    \includegraphics[width=0.5\linewidth]{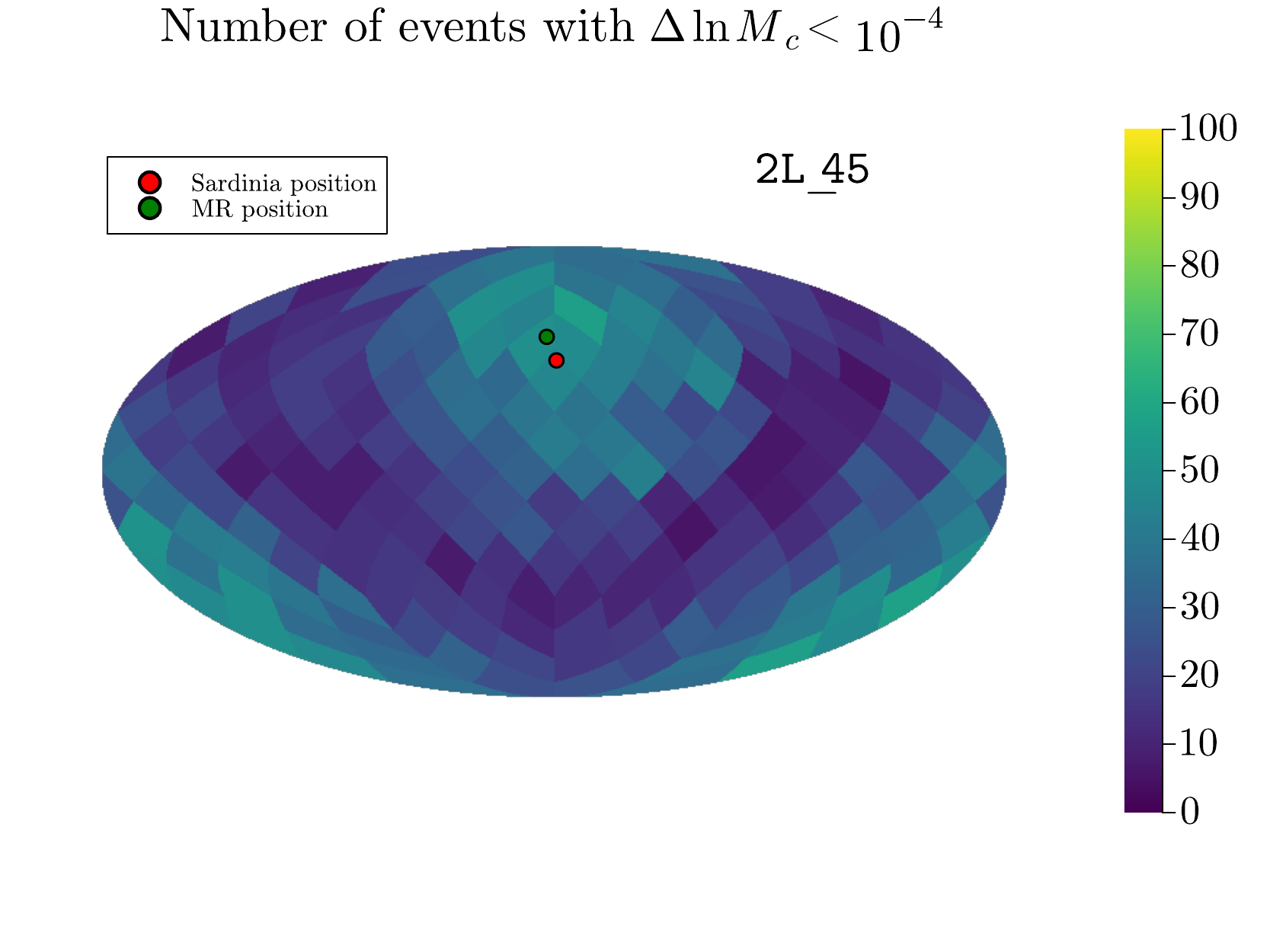}
    \includegraphics[width=0.49\linewidth]{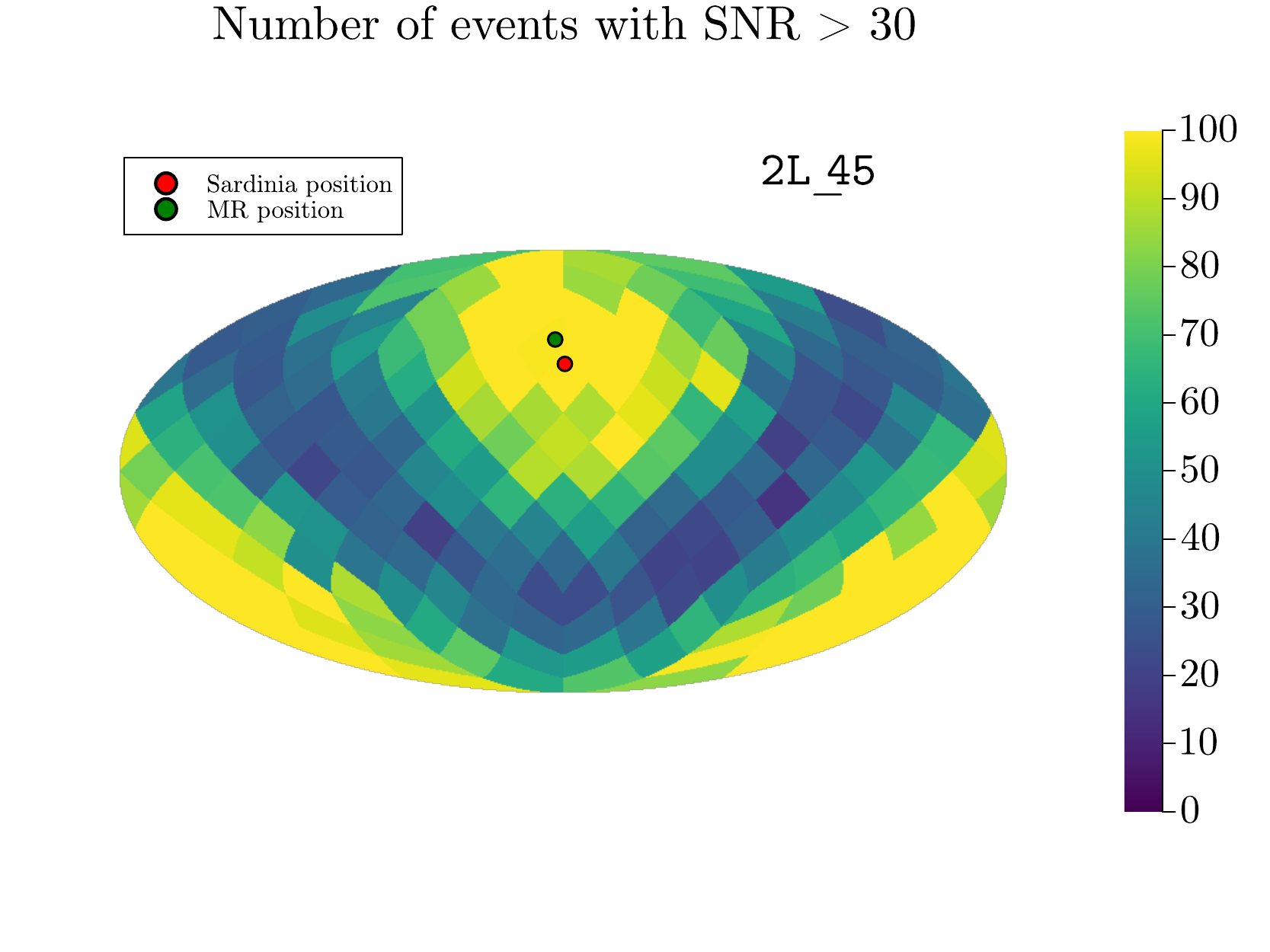} 
    \caption{In Figure, we plot the sky maps as seen by the detectors, and the color coding represents the number of events with a precision better than the threshold indicated in the title. The first column shows the number of events with a relative error better than  $10^{-4}$ in the chirp mass at detector, while the second column shows the number of events with a SNR larger than 30. The first row shows the triangular network, the second row shows the \texttt{2L\_0} network, and the third row shows the \texttt{2L\_45} network. The red and green points represent the two detectors of each network.}
    \label{fig:BBH_sky_maps_app}
\end{figure}

In \cref{fig:BBH_sky_maps_app}, we plot the analogous figures to the ones in \cref{fig:BBH_sky_maps} for the errors on the relative chirp mass and the SNR of the events.
We see that, for the \texttt{2L\_0} network, the SNR seems to contribute to the better angular resolution since we see the same pattern as the one highlighted in \cref{subsec:golden}. Moreover, these patterns follow the ones for the luminosity distance, which match the pattern functions, as one would expect. Instead, the patterns of the sky area are the result of a more complicated concurrence of causes, as described in \cref{subsec:golden}.

\bibliographystyle{JHEP}
\bibliography{bibliography.bib}
\end{document}